\documentclass[preprint,showpacs,preprintnumbers,amsmath,aps,pra]{revtex4-1}

\usepackage{amssymb,amsfonts,amstext,graphics,graphicx,subfigure}
\usepackage[colorlinks]{hyperref}
\usepackage{color}

\usepackage{tikz}
\usepackage{stackengine}

\usepackage{mathbbol}

\def\XXint#1#2#3{{\setbox0=\hbox{$#1{#2#3}{\int}$ }
\vcenter{\hbox{$#2#3$ }}\kern-.5\wd0}}

\def\calo{\mathcal{O}}

\def\R{\mathbb{R}}

\def\M{\mathbb{M}}

\def\bq{\begin{equation}}
\def\eq{\end{equation}}
\def\bqy{\begin{eqnarray}}
\def\eqy{\end{eqnarray}}

\def\bal#1\eal{\begin{align}#1\end{align}}

\def\quadd{\quad\quad}

\def\al{\alpha}
\def\be{\beta}

\def\de{\delta}

\def\ep{\epsilon}

\def\ga{\gamma}
\def\Ga{\Gamma}

\def\ka{\kappa}
\def\la{\lambda}
\def\La{\Lambda}

\def\om{\omega}

\def\si{\sigma}

\def\varep{\varepsilon}

\def\bfV{\mathbf{V}}

\def\p{\partial}

\def\andq{\quad\mathrm{and}\quad}
\def\andqq{\quadd\mathrm{and}\quadd}

\def\ncr{\nonumber\\}

\begin{document}

\title{A Hamiltonian description of  finite-time singularity in Euler's fluid equations}
\author{Philip J.~Morrison}
\email{morrison@physics.utexas.edu}
\affiliation{Department of Physics and Institute for Fusion Studies, 
The University of Texas at Austin, Austin, TX, 78712, USA}
\author{Yoshifumi Kimura}
\affiliation{Graduate School of Mathematics, Nagoya University, Furo-cho, Chikusa-ku, Nagoya 464-8602, Japan}
\date{\today}

\begin{abstract}

The recently proposed low degree-of-freedom model of Moffat and Kimura \cite{MK19,MK19b} for describing the approach to finite-time  singularity of the incompressible Euler fluid equations is investigated.  The  model assumes an initial finite-energy configuration of two vortex rings  placed symmetrically on two tilted planes. The Hamiltonian  structure  of the inviscid limit of the model is obtained. The associated  noncanonical Poisson bracket \cite{pjm98}  and two invariants, one that serves as the Hamiltonian and the other a Casimir invariant,  are discovered.  It is shown that the system is integrable with a solution that lies on the intersection for the two invariants, just as for the free rigid body of mechanics whose solution  lies on the intersection of the kinetic energy and angular momentum surfaces.  Also, a direct quadrature is  given and used to  demonstrate the Leray form for finite-time singularity in the model. To the extent the Moffat and Kimura  model accurately represents  Euler's ideal fluid equations of motion, we have shown the existence of  finite-time singularity.

\end{abstract}

\maketitle

\tableofcontents


\section{Introduction}
\label{sec:intro}

One approach to the pedigreed quest for determining the existence or nonexistence of finite-time singularity in the Euler and Navier-Stokes equations of fluid dynamics (see, e.g., \cite{doering} for an overview)  is to analyze them with very specialized  initial conditions.  Because  the vorticity at a singularity must diverge \cite{B_etal94},   effort has been spent on understanding the behavior of interacting localized de-singularized vortex tubes, leading to the study of vortex reconnection and its role in turbulence.    Various configurations have been proposed and investigated numerically for reconnection in classical turbulence (e.g., \cite{Kt94,K05,HD11}), quantum turbulence (e.g.,  \cite{B_etal08,Z_etal12,VPK17}),  and for the existence of singularity (e.g., \cite{SP85,PGGB,BMH16,KM18,K18}).

The present work investigates the reduced model of Moffatt and Kimura  \cite{MK19,MK19b} that describes the interaction of two circular vortex rings.  To the extent this model accurately represents  Euler's ideal fluid equations of motion, we have shown the existence of  finite-time singularity.  Explicitly, we have shown that within the proposed limits of applicability of this model there exist solutions that blow up in finite time. 

Section \ref{sec:eom} describes the Moffatt and Kimura (MK) model.  This is followed by Sec.\ \ref{sec:HS} where  the Hamiltonian structure of the MK model is given, which is essential for our analysis.  Here we discover two constants of motion for the  MK model. One invariant  serves as  the  Hamiltonian for its noncanonical Hamiltonian formulation (flow on a Poisson manifold; see  \cite{pjm98}), while  the other turns out to be a Casimir invariant.  The Hamiltonian formulation allows us, in Sec.\ \ref{sec:quad},  to obtain geometrical intuition about the solution space by examining the intersection of the level sets of the two invariants, akin to the visualization afforded by the constancy of the energy and angular momentum magnitude for the Euler equations that describe the free rigid body. 
Also in  this section we show how to reduce the MK system to quadrature and obtain for special initial conditions explicit solutions that have exact Leray scaling, which is representative of the finite-time singularity. Conditions for singularity within the range of applicability of the derivation given  in Refs.~\cite{MK19,MK19b} are presented. 
In Sec.\ \ref{sec:conclu} we summarize  our results and mention some  future avenues.  Appendices are provided that exhibit additional features of our results.

\section{The Moffatt-Kimura system}
\label{sec:eom}

The MK system is a three-dimensional system of ordinary differential equations that describes the evolution of two initially circular vortices of radius $R$ and circulations $\pm\Ga$, located symmetrically on planes $x=\pm z\tan \al$, with pitch angle $\al$,  in an $(x,y,z)$ Cartesian coordinate system.  The system is written in dimensionless form using the space scale $R$,  time scale  $R^2/\Ga$, and effective Reynolds number $R_\Ga= \Ga/\nu=: 1/\varep >> 1$.  Here $\nu$  is the usual
kinematic viscosity of the fluid. It is assumed that the vortices have Gaussian cores of radius measured by  $\de$, a separation measured by $s$, and a curvature given by $\ka$,  which evolve in terms of the dimensionless time according to 
\bal
\dot{(\de^2)}&=\varep - c_2 \frac{\ka \de^2}{s}\,,
\label{dotde0}\\
\dot{s}&=-c_2\ka \left[\ln\left(\frac{s}{\de}\right) +\be_1\right]
\label{dots0}\\
\dot{\ka}&=c_1\frac{\ka}{s^2}\,,
\label{dotka0}
\eal
subject to the  inequality constraints  for applicability
\bq
\de< s< 1/{\ka}\,,
\label{ineq}
\eq 
which are assumed in the derivation.  Roughly speaking, these inequalities assure that the vortex cores, with size measured by $\de$,  are small enough so that cores do not overlap as the rings merge, i.e., as their separation given  by $s$ decreases, and that portions of the rings are sufficiently far away, as measured by $1/\kappa$,  to allow far field expansion. 
The constants appearing in Eqs.~\eqref{dotde0}, \eqref{dots0}, and \eqref{dotka0}    are given by  
\bq
c_1=  \frac{\cos\al \sin\al}{4\pi} \qquad\mathrm{and}\qquad 
c_2=  \frac{\cos\al }{4\pi} \,,  
\label{alpha}
\eq 
and the parameter $\be_1$ depends on the vortex core,  with  the value  $\be_1=0.4417$ for a gaussian core profile. 
  
If the MK system is to be a  reduction that  inherits the noncanonical Hamiltonian structure of Euler's fluid equations (see e.g.\ \cite{pjm98}) then it will have a Hamiltonian form upon setting the  $\varep=0$.  Thus we investigate 
 \bqy
   \dot{\de}&=&-\frac{c_2}{2}\,   \frac{\ka \de }{s}
  \label{dotde}\\
 \dot{s}&=&- c_2\,  \ka \left[ \ln\left(\frac{s}{\de}\right) +\be_1\right]
   \label{dots}\\
 \dot{\ka}&=&c_1\,  \frac{\ka}{s^2}\,.
  \label{dotka}
\eqy 
For  later use we record here the equation of motion for $x:=s/\de$, 
\bq
\dot{x}= - \frac{c_2 \ka}{\de}(\La -1/2)\,,
\label{dotx}
\eq
which is easily verified.  Here, for convenience,  we have defined
 \bq
 \La := \ln\left(\frac{s}{\de}\right) +\be_1 = \ln(x) +\be_1 \,.
 \label{La}
 \eq
  The smallest value of $\La$ consistent with inequality \eqref{ineq} occurs when $\de=s$; thus, 
  \bq
 \La    \geq  \be_1 \qquad \mathrm{or}\qquad  \La -1/2 \geq \be_1-1/2 = -0.0583 \,,
 \label{critLa}
 \eq
 where the gaussian value of $\be_1=0.4417$ is used in the  second expression.  We note here that 
 \bq
 \La =1/2\qquad \mathrm{ at}\qquad  x^*= e^{1/2-\be_1}\approx 1.0600\,.
 \eq
 
\section{Hamiltonian Structure}
\label{sec:HS}

\subsection{Generalities}

Hamiltonian systems are usually written in terms of canonically conjugate sets of variables, a configuration space coordinate and its conjugate momentum.  The noncanonical Hamiltonian description is one where the form in terms of canonical variables is not necessary and replaced by algebraic properties of the Poisson bracket.   The terminology noncanonical Hamiltonian was introduced in the context of the ideal fluid and magnetohydrodynamics in  \cite{pjmG80}, but the ideas date back to the work of Sophus Lie.  (See, e.g.,  \cite{s&m,pjm98,pjmAP20} for review.)

Given an $n$-dimensional  phase space with  coordinates $z=(z^1, z^2,\dots, z^n)$, a system of ordinary differential equations has noncanonical  Hamiltonian form  if there exists a Poisson bivector $J$,  an antisymmetric  second rank contravariant  tensor, 
and a conserved phase space function $H(z)$ such that  the equations  can be written as follows:
\bq
\dot{z}^i = \{z^i,H\}= J^{ij} \frac{\p H}{\p z^j}  \qquad i,j=1,2,\dots n\,,
\label{genHamform}
\eq
where repeated indices are summed and the Poisson bracket defined on functions of the coordinate $z$, 
\bq
\{f,g\}= \frac{\p f}{\p z^i} J^{ij}  \frac{\p g}{\p z^i} \,
\eq
is bilinear, antisymmetric, and most importantly satisfies the Jacobi identity, 
\bq
\{f,\{g,h\}\} +\{f,\{g,h\}\} +\{f,\{g,h\}\} =0\,,
\label{jacobi0}
\eq  
for all functions $f,g,h$.  Unlike for the canonical description,  the tensor $J$,  the Poisson tensor, may depend on the coordinate $z$. It generates the Hamiltonian vector field as depicted on the righthand side of the second equality of \eqref{genHamform}. In  coordinates, \eqref{jacobi0} is equivalent to  the vanishing of  the following purely antisymmetric three-tensor:
\bq	
S^{ijk}= J^{i\ell}\frac{\p J^{jk}}{\p z^\ell} + J^{j\ell}\frac{\p J^{ki}}{\p z^\ell} + J^{k\ell}\frac{\p J^{ij}}{\p z^\ell}  \equiv 0\,, 
\label{jacobi0c}
\eq
a quantity that is checked in practice. 

When $\det J\neq 0$, an old theorem of Darboux, based on the algebraic properties of the Poisson bracket, implies that there is a coordinate change from the noncanonical coordinates $z$ to a set of canonically conjugate coordinates.  However, when $\det J=0$ this is not possible because of degeneracy, i.e., the existence of special functions $C$, called Casimir invariants,  that satisfy $\{C,f\}=0$ for all phase space functions $f$.   Thus, Casimir invariants are built-in to the phase space, for they will be conserved by a system generated by any Hamiltonian function. In the coordinates $z$, $C$ is a Casimir invariant if it satisfies
\bq
J^{ij}\frac{\p C}{\p z^j}=0\,. 
\label{cascond}
\eq

Noncanonical Hamiltonian systems possess the rich geometrical structure of so-called Poisson manifolds, where through every point of the phase space manifold is a conserved canonical Hamiltonian subspace, i.e., the manifold is foliated by symplectic leaves (see, e.g., the seminal reference \cite{weinstein} and the recent contribution \cite{pjmY20}).   We will see in practical terms how an interesting  Poisson  manifold emerges from the MK system.

For three-dimensional systems, like the MK system,  the Poisson tensor  $J$ has the form
\bq
J = 
 \begin{bmatrix} 
0& V_3& -V_2\\
-V_3& 0&  V_1\\
V_2& - V_1& 0\\
\end{bmatrix}
\eq
for some vector $\bfV(z)=(V_1,V_2,V_3)$,  and it can be  shown easily that \eqref{jacobi0},  for the Jacobi identity,  is satisfied if  $\bfV(z)$ satisfies 
\bq
\bfV\cdot \nabla \times \bfV =0\,.
\label{jacobi1}
\eq
Thus, for three-dimensional systems there is a  convenient way to check the Jacobi identity.  Because antisymmetric matrices have even rank, $J$ must have rank 2 or 0.  The latter of course would generate trivial dynamics -- of interest is the case of rank 2 where there is a single Casimir invariant, and condition \eqref{cascond} can be written compactly as
\bq
\bfV\times \nabla C=0\,.
\label{3dcas}
\eq
Relations \eqref{jacobi1}  and \eqref{3dcas} play central roles in our discovery of the Hamiltonian structure of the MK system.

\subsection{The Hamiltonian and Poisson bracket}
\label{ssec:mkham}

Given the equations of motion of a system, like \eqref{dotde}, \eqref{dots}, and \eqref{dotka}, and a constant of motion, one can seek a Poisson tensor by matching to the equations of motion while enforcing the Jacobi identity.   Thus we seek a suitable invariant, one that physically we expect to be an energy-like quantity.   Using the structure of the MK system  and some insight we find   \eqref{dotde}, \eqref{dots}, and \eqref{dotka} conserve the following:
 \bq
H=\frac1{\de^2}\left[\ln\left(\frac{s}{\de}\right) +\be_1-\frac12\right]= \frac1{\de^2}\big[ \La - 1/2\big]\,, 
 \label{H}
\eq
 which can be shown directly. From \eqref{H} we obtain
 \bq
 s=\de\, e^{\de^2 {H} -\be_1 +1/2}\,.
 \label{sde}
 \eq
 Again, as with \eqref{critLa}, the threshold  for inequality \eqref{ineq} occurs when $\de=s$, yielding
  \bq
 H > (\be_1-1/2)/\de^2=-0.0583/\de^2\,,
 \label{critH}
 \eq
 where again the gaussian value of $\be_1$ is used in the  equality. 

In terms of the  coordinates $z=(\de, s, \ka)$,  the analog of equation \eqref{genHamform} for the MK system takes the form
\bq
 \begin{bmatrix} 
\dot{\de}  \\
\dot{s} \\
\dot{\ka}\\
\end{bmatrix}
= 
 \begin{bmatrix} 
0& V_3& -V_2\\
-V_3& 0&  V_1\\
V_2& - V_1& 0\\
\end{bmatrix}
 \begin{bmatrix} 
\p H/\p \de \\
\p H/\p s \\
\p H/\p \ka \\
\end{bmatrix}\,,
\label{form}
\eq
which upon making use of the MK equations of motion  \eqref{dotde}, \eqref{dots}, and \eqref{dotka}, and the candidate Hamiltonian \eqref{H}, we obtain the following equation for the Poisson tensor:
\bq
 \begin{bmatrix} 
-{c_2}\, {\de \ka}/{(2s)} \\
- c_2\,  \ka \, \La  \\
c_1\,  {\ka}/{s^2}\\
\end{bmatrix}
= 
 \begin{bmatrix} 
0& V_3& -V_2\\
-V_3& 0&  V_1\\
V_2& - V_1& 0\\
\end{bmatrix}
 \begin{bmatrix} 
-2 \La  /\de^3\\
1/( s \de^2)\\
0 \\
\end{bmatrix}
\label{form20}\,.
\eq
Thus, the goal is to solve \eqref{form20} for a $\bfV=(V_1,V_2,V_3)$ that satisfies \eqref{jacobi1}.

It follows immediately from \eqref{form20} that
\bq
\dot{\de}=-{c_2}\, {\de \ka}/{(2s)} =V_3/(s\de^2)
\andqq
\dot{s}= - c_2\,  \ka  \La   = 2 V_3  \La  /\de^3\,;
\label{V3a}
\eq
thus, 
\bq
V_3= -c_2 \ka \de^3/2
\label{V3}
\eq
works for both equations  of \eqref{V3a}.  The remaining equation yields the expression, 
\bq
\dot{\ka}= c_1\ka/s^2= -2 V_2 \La  /\de^3  - V_1/(s\de^2)\,.
\eq
Thus, it appears there is  freedom in the choices of $V_1$ and $V_2$ to satisfy \eqref{jacobi1}, 
the Jacobi identity.  Upon setting
\bq
V_1 = -c_1\ka\de^2 /s -2 V_2 \La  s/\de \,.
\label{V1}
\eq
we seek to find a $V_2$ that ensures   \eqref{jacobi1} is satisfied.  A direct calculation implies 
 \bal
\bfV\cdot\nabla\times \bfV 
&= - \frac{c_2}{2}\,  \ka \de^3 \, \frac{\p  V_2 }{\p \de}
-  c_2\,  \ka s\de^2 \La   \,  \frac{\p V_2 }{\p s}
+  c_1\,  \frac{\ka\de^2}{s}  \, \frac{\p  V_2}{\p \ka}
\nonumber\\
& + V_2\left(-c_1\, \frac{\de^2}{s} + \frac{c_2}{2}\, \ka \de^2 
-c_2\,  \ka \de^2   \La  
\right)
+   \frac{c_1 c_2}{2}\, \frac{\de^5 \ka^2}{s^2}\,.
\label{jacobi2}
\eal
and the goal is to find a $V_2$ such that \eqref{jacobi2} vanishes.

Upon  inserting the following into \eqref{jacobi2}, 
\bq
V_2= \ka\de A(x)\,,
\label{V2}
\eq
where recall $x:=s/\de$ and  using 
\bq
\p_k V_2= \de A(x)\,, \qquad \p_s V_2 = \ka A'(x)\,, 
  \qquad \p_\de V_2= \ka  A(x) - \ka x A'(x)
\eq
\eqref{jacobi2} becomes
\bq
\bfV\cdot\nabla\times\bfV=-c_2\ka^2 \de^3\left[x A' \left(\La -\frac12\right) + \La A -\frac{c_1}{2x^2}\right]\,.
\eq
Thus,  the Jacobi identity is satisfied provided  we can find a solution $A(x)$ to  
\bq
x A' \left(\La -\frac12\right) + \La A -\frac{c_1}{2x^2}=0\,,
\label{Aeq}
\eq
where recall $\La (x)$ is given by \eqref{La}.   If such a function $A$ is found, then the MK system is Hamiltonian with a bracket defined by 
\bq
V_1= -c_1\frac{\ka\de^2}{s} -2 \ka \La  s \, A(x)  \,,\quad 
V_2=\ka\de \,A(x)\,,
\andq
V_3= -\frac{c_2}{2} \ka \de^3\,.
\label{Vs}
\eq

Thus, we proceed to solve for $A$.   In the light of the inequality \eqref{ineq},   $\La $ is seen to be a positive monotonic function with the  inverse
\bq
x=e^{\La -\be_1}\,.
\label{Lainv}
\eq
Consequently, we can use   $\La $ as the independent variable and  rewrite \eqref{Aeq} as 
\bq
\frac{d}{d\La } \left( e^\La \sqrt{\La -1/2\,}\, A\right) = \frac{c_1}{2x^2} \frac{e^\La }{\sqrt{\La -1/2\,}}
 =\frac{c_1e^{2\be_1}}{2} \frac{e^{-\La }}{\sqrt{\La -1/2\,}}\,.
 \label{Lageq12}
 \eq
 Here we have assumed $\La -1/2\geq 0$.  In light of  \eqref{critLa} this may not be true; thus, we  will return and consider the case where $\La -1/2\leq 0$.

 Antidifferentiating both sides gives
\bq
A=\frac{e^{-\La }}{\sqrt{\La -1/2\,}} \frac{c_1e^{2\be_1-1/2}}{2 }  \int^{\La -1/2}  \frac{e^{-\ell}}{\sqrt{\ell\,} }\, \,d\ell\,.
\eq
Then, with the definition of  the incomplete gamma function  
\bq
\ga(\si,u)=\int_0^u  u'^{\si-1} e^{-u'} du'\,, 
\label{ligf} 
\eq
and the identity
\bq
\ga(1/2,u)=\sqrt{\pi} \mathrm{erf}(\sqrt{u})=2 \int_0^{\sqrt{u}} e^{-u'^2} d u'\,,
\label{gamaerf}
\eq
with erf being the error function, we obtain

\bq
A=c_0\,  \frac{e^{-\La }}{\sqrt{\La -1/2\,}}    
+  \frac{\sqrt{\pi} c_{1} \,  }{2}\,  e^{2\be_1-1/2} \, \frac{e^{-\La }}{\sqrt{\La -1/2\,}}\,  \mathrm{erf}\left(\sqrt{\La -1/2}\right)  
\,,
 \label{A1}
\eq
where $c_0\in\R$ is the integration constant.   Upon inserting \eqref{A1} into the expressions of  \eqref{Vs}, we have a  one-parameter family of Poisson brackets. 
It is convenient to choose $c_0=0$,  giving
\bq
A=\frac{\sqrt{\pi} c_{1} \,  }{2\, x}\, \frac{e^{\be_1-1/2} }{\sqrt{\La -1/2\,}}\,  \mathrm{erf}\left(\sqrt{\La -1/2}\right)
\,,
\label{Af}
\eq
which is valid for $\La -1/2\geq 0$, but we will see it is also valid for $\La -1/2<0$. Observe, the choice $c_0=0$  gives us regularity at $\La =1/2$.   Inserting \eqref{Af} into the equations of \eqref{Vs} defines  the  noncanonical Poisson bracket that we will use  for $\La  \geq 1/2$.  

Now consider the case where  $\La  \leq 1/2$, which occurs for 
\bq
0\leq 1/2- \La \leq 1/2-\be_1 - \ln x\leq 1/2-\be_1 \,,
\label{Larange}
\eq
where we assume $x\geq 1$ consistent again with \eqref{ineq}.  Instead of \eqref{Lageq12}, consider
\bq
\frac{d}{d\La } \left( e^\La \sqrt{1/2-\La \,}\, A\right) = - \frac{c_1 e^{2\be_1}}{2} \,\frac{e^{-\La }}{\sqrt{1/2-\La } }\,,
 \label{L leq12}
 \eq
where \eqref{Aeq} has been used.   Integrating both sides   from $\La =1/2$  leads to 
  \bq
 A=   {c_1 \,e^{2\be_1 -1/2}\, e^{-\La }}  \, \,  M(1/2,3/2,1/2-\La )\,, \qquad  \La  \leq 1/2\,,
 \label{Af2}
 \eq
 where 
 \bq
 M(1/2,3/2,z)= \frac12\int _0^1 \frac{e^{zt}}{\sqrt{t}}\, dt 
 = \frac{1}{2\sqrt{z}}\int _0^z \frac{e^{u}}{\sqrt{u}}\, du\,,
 \eq
is the   Kummer function \cite{a&s},  which is  sometimes  called the confluent hypergeometric function of the first kind and denoted by ${}_1F_1(1/2;3/2;z)$.  Inserting \eqref{Af2} into the equations of \eqref{Vs} defines  the  noncanonical Poisson bracket defined  for $\be_1\leq \La  \leq 1/2$.

A comparison of \eqref{Af} and \eqref{Af2} follows from the identity
\bq
\mathrm{erf}(z)= \frac{2z}{\sqrt{\pi}}\,M(1/2,3/2,-z^2)\,,
\eq
which implies 
\bq
\frac{\sqrt{\pi}}{2}\frac{1}{\sqrt{\La -1/2}}\, \mathrm{erf}\left(\sqrt{\La -1/2}\right) = M(1/2,3/2,1/2-\La ))\,.
\eq
Because  $\mathrm{erf}(ix)$ for $x\in\R$  is pure imaginary,  like $\sqrt{\La -1/2}$ for $\La <1/2$, we can analytically continue the expression in terms of the error function through zero  to pure imaginary values yielding a real quantity.  Thus, expression \eqref{Af} can be used for both  $\La \geq 0$ and $\La <0$.

\subsection{The Casimir Invariant}
\label{ssec:cas}

\begin{figure}[htb]
\includegraphics[scale=.5]{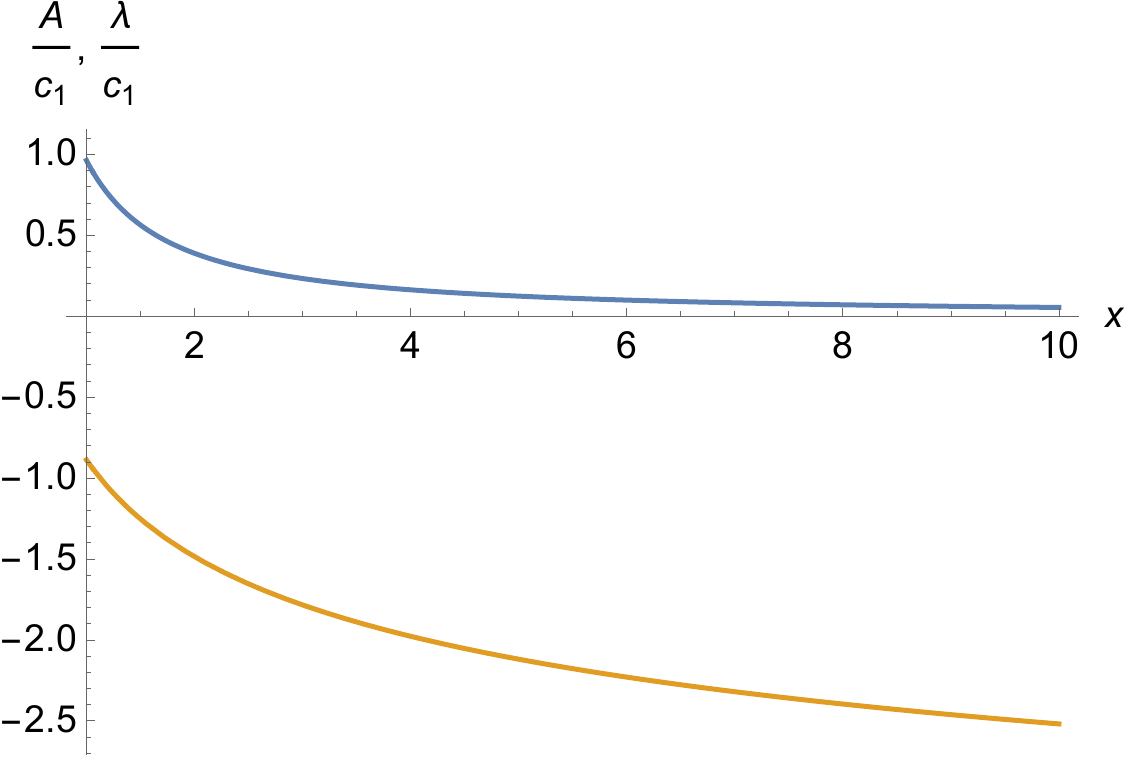}
\caption[sm]{Plots of $A/c_1$ (blue) and  $\la/c_1$ (orange) vs.\ $x=s/\de$.  Recall $\la'=-A(x)$ [cf.\ Eq.~\eqref{laprime}].}
\label{fig:Alambda}
\end{figure}

Given the Hamiltonian structure of Sec.~\ref{ssec:mkham} we know immediately that the MK system  must possess  another constant of motion, the Casimir invariant.  A little thought reveals 
the  Casimir must have the form
\bq
C= \frac{c_2}{2} \ka + \frac{\la(x)}{\de}\,.
\label{C}
\eq
In the present context,  \eqref{3dcas} is equivalent to 
\bq
 \begin{bmatrix} 
0& V_3& -V_2\\
-V_3& 0&  V_1\\
V_2& - V_1& 0\\
\end{bmatrix}
 \begin{bmatrix} 
\p C/\p \de \\
\p C/\p s \\
\p C/\p \ka \\
\end{bmatrix}
=
\begin{bmatrix} 
0& V_3& -V_2\\
-V_3& 0&  V_1\\
V_2& - V_1& 0\\
\end{bmatrix}
 \begin{bmatrix} 
-\la/\de^2-s\la'/\de^3 \\
\la'/\de^2\\
c_2/2 \\
\end{bmatrix}
=0
\label{cas}
\eq
The first equation  gives 
\bq
0= -\frac{c_2}{2} \ka  {\de^3}\frac{\la'}{\de^2} - \ka\de A \frac{c_2}{2}
\quad\Rightarrow\quad\la'=-A\,,
\eq
while the second gives after some manipulation
\bq 
0=-\frac{c_2\ka\de}{2}\left(\la - xA +\frac{c_1}{x}+2Ax\La 
\right)\,.
\eq
Because of the Jacobi identity,  the third equation must automatically  be solved by the above, which can be  verified directly.  
Thus we have
\bal
\la'(x)&=-A(x)
\label{laprime}
\\
\la(x)&= Ax(1-2\La )-\frac{c_1}{x}=-2Ax(\La -1/2) - \frac{c_1}{x} \,,
\label{lam}
\eal
where $A$ solves \eqref{Aeq}.   Upon differentiating \eqref{lam} we see that if $A$ satisfies \eqref{Aeq}, then \eqref{laprime} is automatic for a solution of   \eqref{lam}. In particular, using \eqref{Af} we obtain
\bq
 \la(x) = -  \sqrt{\pi} c_{1} \, e^{\be_1-1/2} \, {\sqrt{\La -1/2\,}}\,  \mathrm{erf}\left(\sqrt{\La -1/2}\right)   
 - \frac{c_1}{x} \,,
 \label{laf}
\eq
where use has been made of  $xe^{-\La }= e^{-\be_1}$. Figure \ref{fig:Alambda} displays plots of both $A/c_1$ of \eqref{Af} (upper)
and $\la/c_1$ \eqref{laf} (lower) as  functions of $x$,  beginning with its critical value of $x=s/\de=1$, revealing that both are  relatively simple monotonic functions consistent with inequality \eqref{ineq}. 

Thus to summarize,  the Casimir, a second invariant, is given by
\bal
C&= \frac{c_2}{2} \ka + \frac{\la(x)}{\de}
\ncr
&=  \frac{c_2}{2} \ka   -  \sqrt{\pi} c_{1} \, e^{\be_1-1/2} \, \frac{1}{\de} {\sqrt{\La -1/2\,}}\,  \mathrm{erf}\left(\sqrt{\La -1/2}\right)   
 - \frac{c_1}{s}\,,
\label{Cfinal}
\eal
where we have inserted  $\la$ from \eqref{laf} in \eqref{Cfinal}  and used $x=s/\de$.  We remind the reader that \eqref{Cfinal} is valid for $\La <1/2$ as well as for $\La \geq1/2$.

Because of inequality \eqref{ineq}, not all values of \eqref{Cfinal} are permissible.  To understand the permissible range, we rewrite  \eqref{Cfinal}  as follows:
\bq
\frac{s}{c_1}C-\frac{c_2}{2c_1}\ka s= F(\La)\,,
\eq
where
\bq
F(\La )=  \sqrt{\pi}   \, e^{\La -1/2} \,  {\sqrt{\La -1/2\,}}\,  \mathrm{erf}\left(\sqrt{\La -1/2}\right)   + 1 >0\,.
\label{FLa }
\eq
As noted in \eqref{critLa},   the smallest allowable value of $\La $ is $\be_1$, and it is not hard to show that $F(\La )$ obtains its minimum value at $\be_1$, at  which it is positive.  Thus we see, the Casimir must satisfy 
\bq
C<\frac{c_2}{2} \ka
\label{Cineq}
\eq
to be consistent with the inequality \eqref{ineq}.

\section{The nature of the solution, reduction to quadrature, and analysis}
\label{sec:quad}

\subsection{Geometrical solution}

Given that we have a three-dimensional system with two constants of motion, the solution space can be visualized by examining the  intersection of the level sets  of the  Hamiltonian $H$ of \eqref{H} with those of the Casimir $C$ of \eqref{Cfinal}.   This is a direct analog of how the stable and unstable trajectories of the  the free rigid body, as governed by Euler's equations,  are understood in terms of  the intersection of the angular momentum spheres with the energy ellipsoids.  The same situation occurs for other noncanonical Hamiltonian systems such as the Kida problem of fluid mechanics \cite{pjmMF97} and  the rattleback toy \cite{pjmYT17}, and indeed a large class of flows on Poisson manifolds \cite{pjmY20}.  Thus, by plotting level sets of  \eqref{H} and \eqref{Cfinal}  the nature of trajectories is revealed and, in addition, one can delineate the accessible phase space consistent with \eqref{ineq}.

\begin{figure}[htb]
\centering
\subfigure[{\footnotesize \  }]{\includegraphics[width=0.43\textwidth]{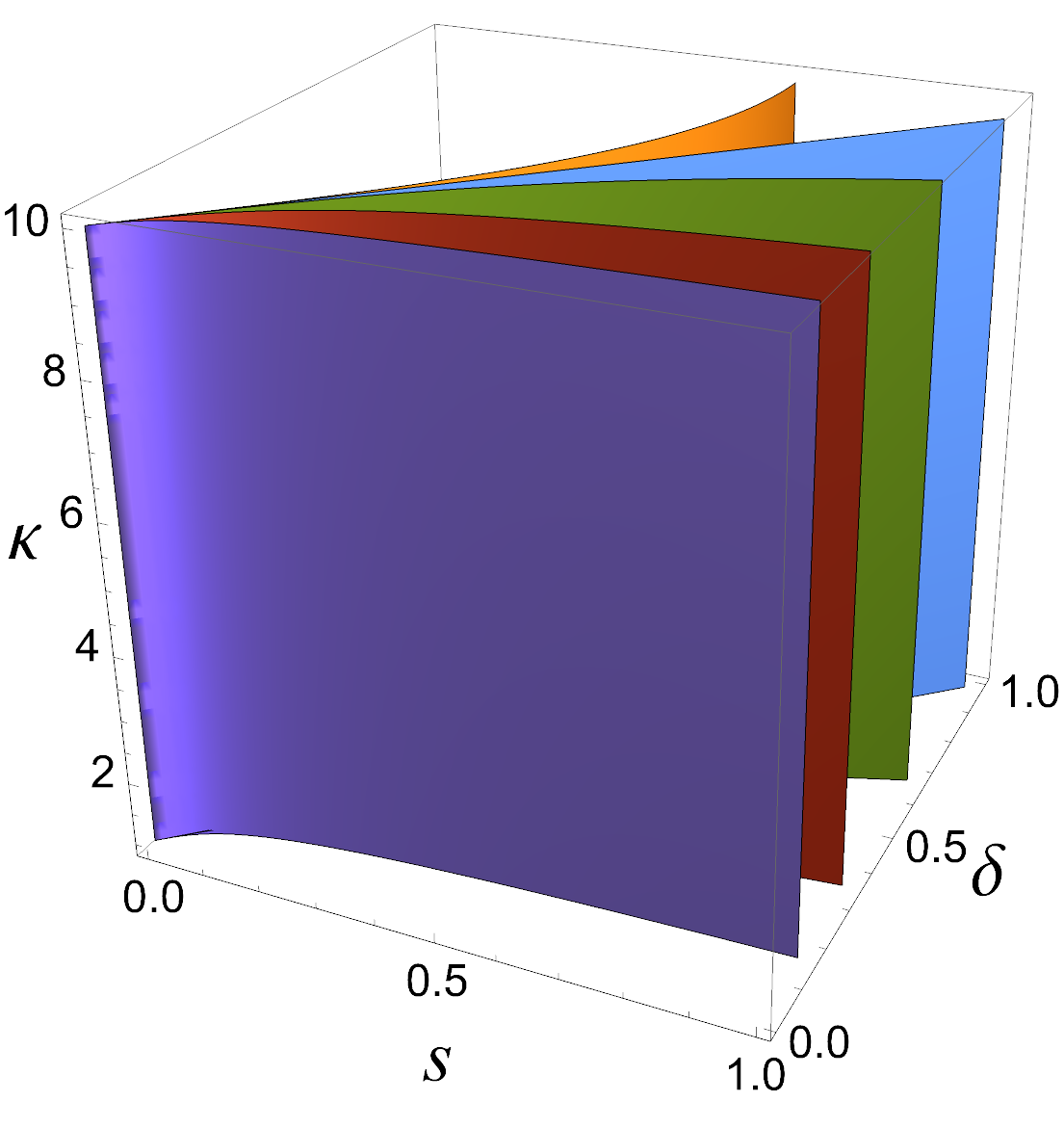}
\label{fig:Hcontoursa}
}
\hspace{.5cm}
\subfigure[{\footnotesize \  }]{
\def\big{\includegraphics[height=5cm]{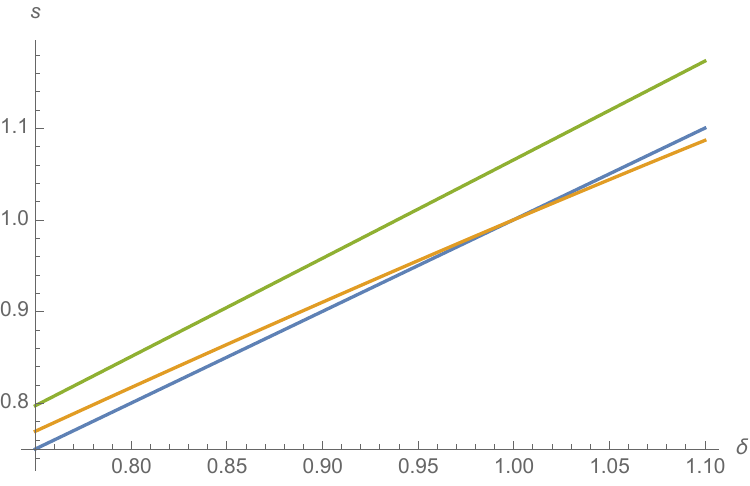}}
\def\little{\includegraphics[height=1.5cm]{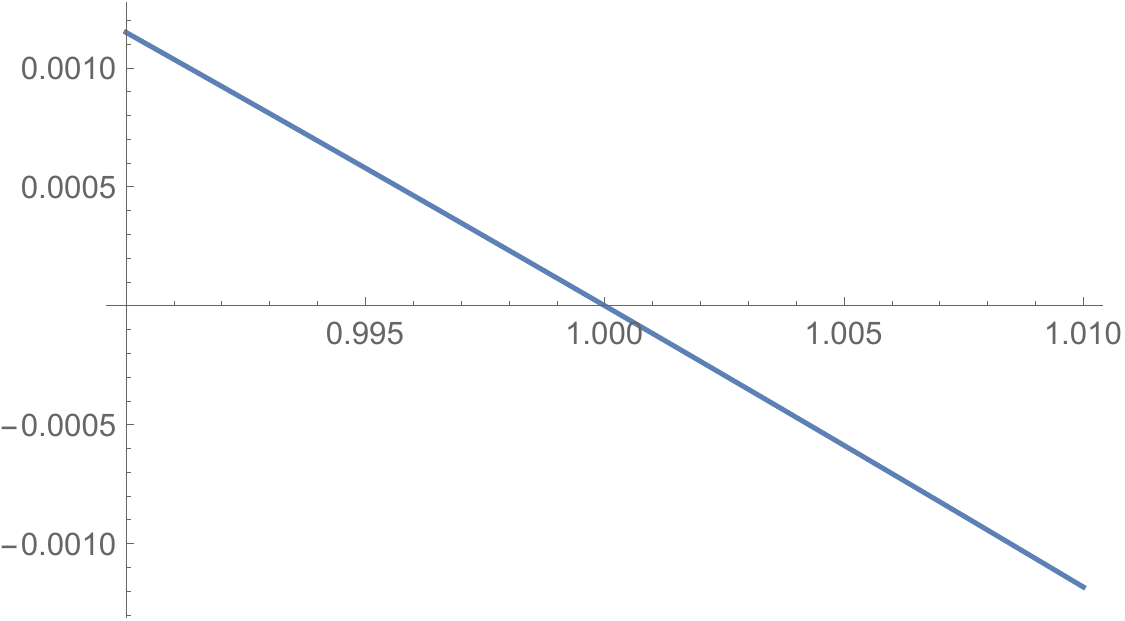}}
\def\stackalignment{r}
\bottominset{\little}{\big}{20pt}{5pt}
\label{fig:Hcontoursb}
}
\caption[sm]
{Contour plots of the Hamiltonian $H$ of \eqref{H}. (a) $\mathring{H}={-.50}$ (orange), $\mathring{H}={0}$ (blue), $\mathring{H}={1}$ (green), $\mathring{H}={10}$ (magenta), and $\mathring{H}={100}$ (lavender).  (b) Closeup plots of $s$ vs.\ $\delta$ at arbitrary $\ka$.  Note the  region to the left of  $\de=1$ for the case where $\mathring{H}=-0.0583$ (orange)  compared to the case $s=\delta$ (blue) with the inset showing the difference  (orange-blue), showing that  $s>\de$ to the left.  The other curve  with   $\mathring{H}=0.005$ (green) is added for comparison.
}
\label{fig:Hcontours}
\end{figure}

 Figure~\ref{fig:Hcontours} displays contours of the Hamiltonian $H$ of \eqref{H}. Since $H$ is independent of the variable $\ka$ it has  translational   symmetry along the $\ka$ axis and, as can be seen in Fig.~\ref{fig:Hcontoursa}, has a sheet-like topology for positive  values of the variables.  Level sets corresponding to different  signs of $H$ have opposite curvature in the $s-\de$ plane,  with there being a region of negative values of $H$ consistent with \eqref{critH} and \eqref{ineq}.   This `negaive energy' interval is shown in Fig.~\ref{fig:Hcontoursb} for $\de<1$.  For all values of $H$,  the sheets become tangent to the $H=0$ plane with a slope given by $s= \de e^{1/2-\be_1}$.   Consequently, if $\de$ and $s$ approach zero, they  do so in a clear and universal way independent of $\mathring{H}$,  the initial  value of the Hamiltonian constant of motion. 
 
\begin{figure}[htb]
\centering
\subfigure[{\footnotesize \  }]{\includegraphics[width=0.45\textwidth]{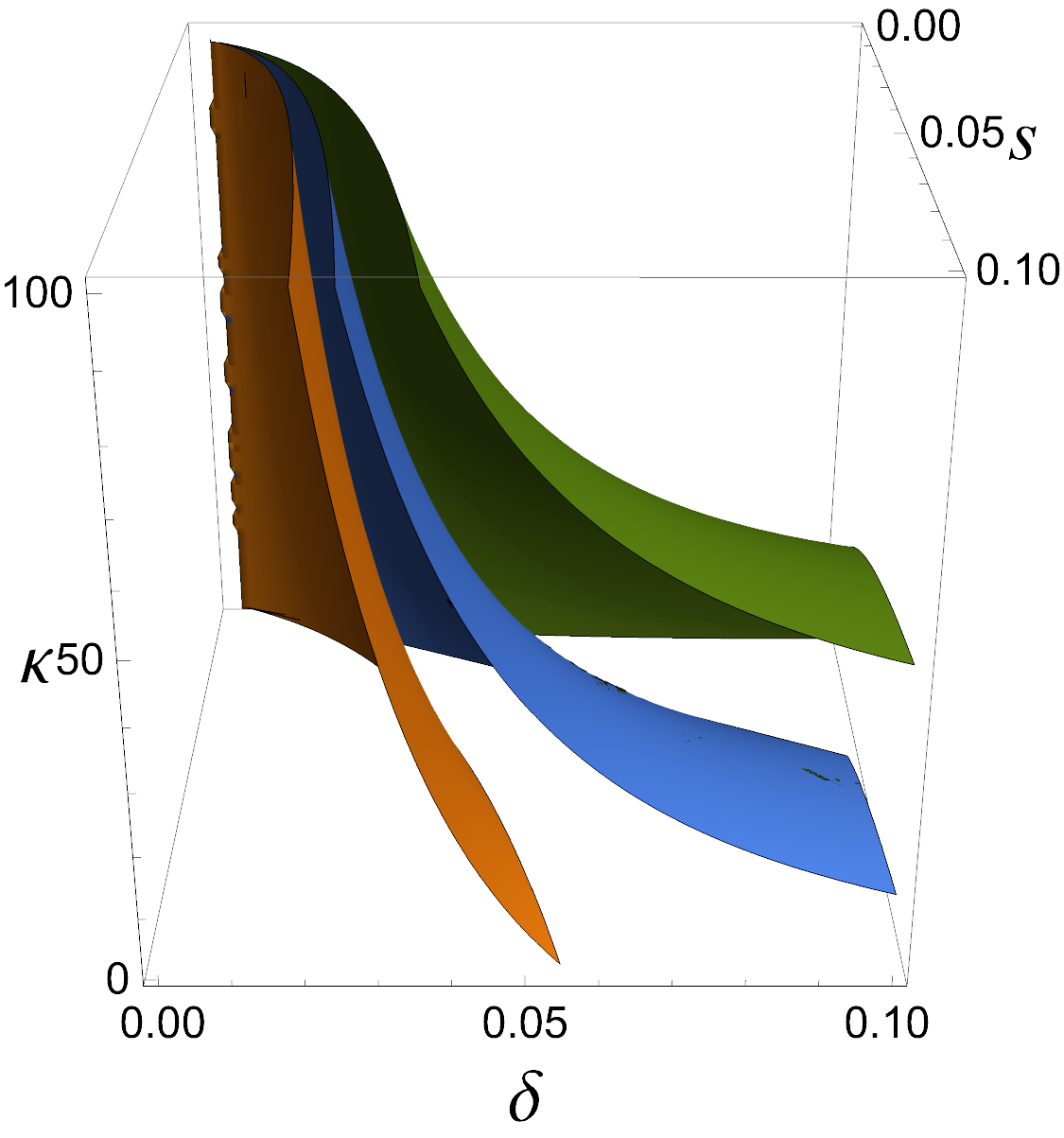}
\label{fig:Ccontours4a}
}
\hspace{.5cm}
\subfigure[{\footnotesize \  }]{\includegraphics[width=0.45\textwidth]{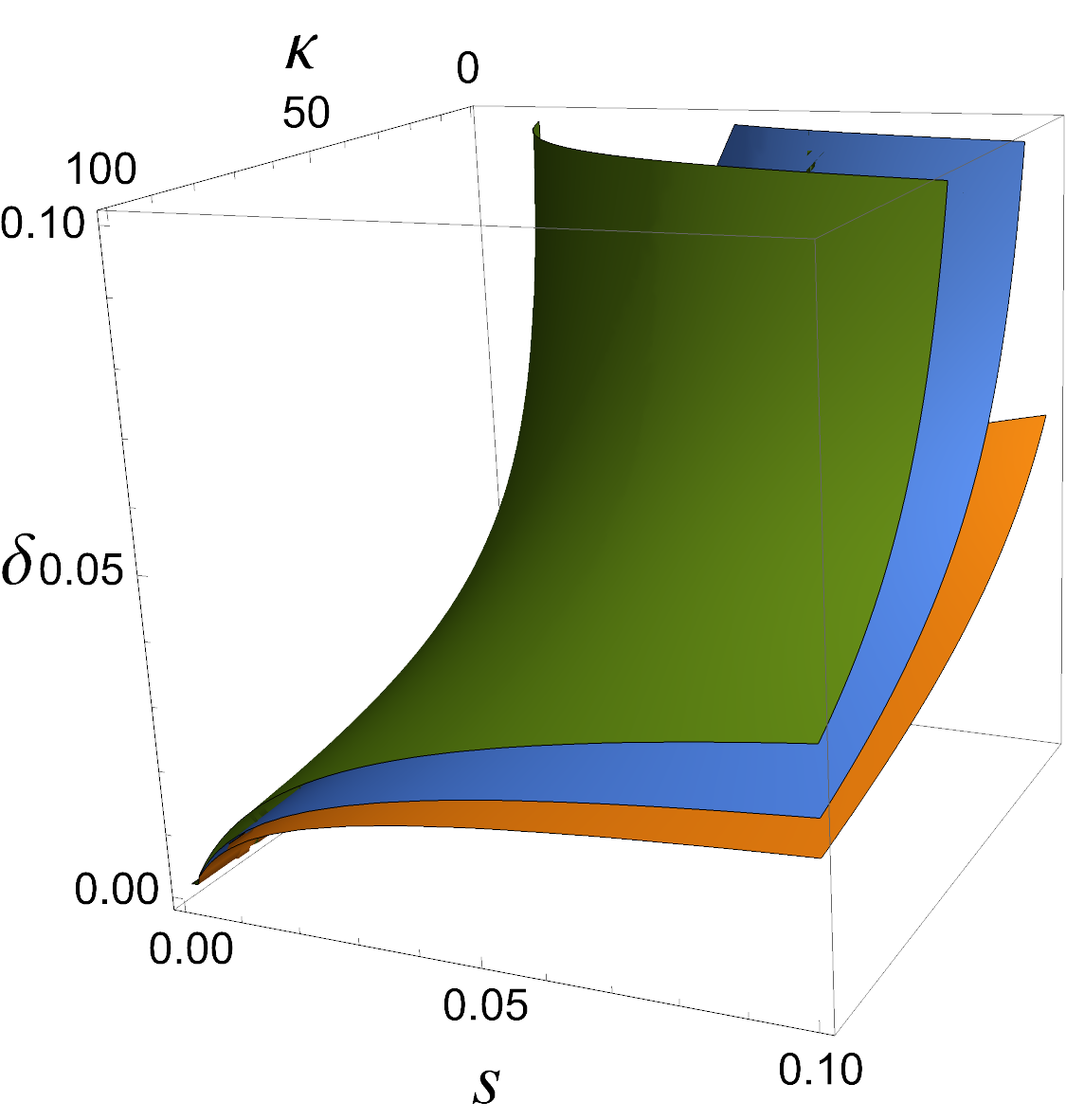}
\label{fig:Ccontours4b}
}
\caption[sm]
{(a) View of the level sets of the  Casimir $C$ for $\al=\pi/4$. (b)    Same as (a) but rotated.
}
\label{fig:Ccontours4}
\end{figure}

\begin{figure}[htb]
\centering
\subfigure[{\footnotesize \  }]{\includegraphics[width=0.44\textwidth]{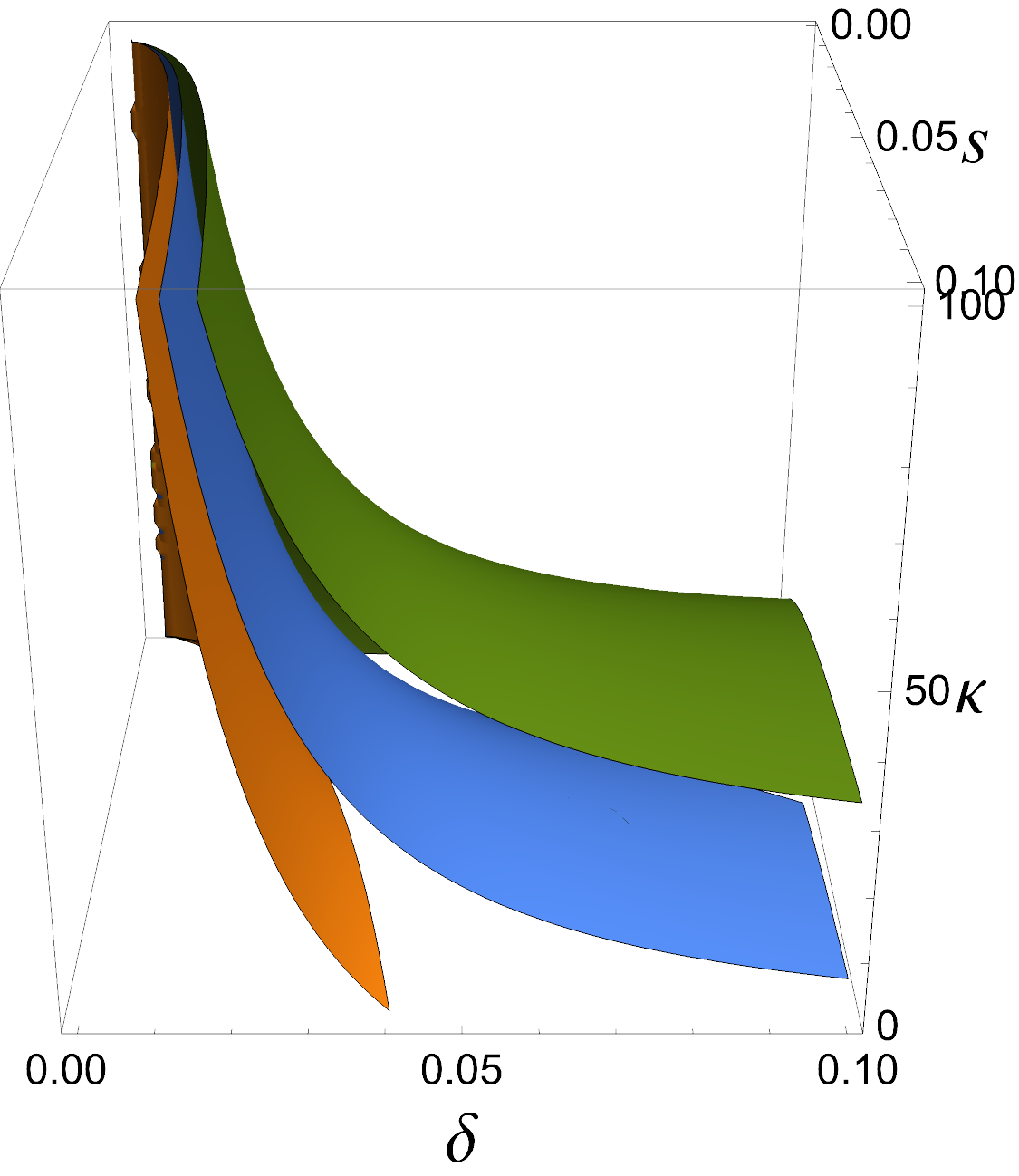}
\label{fig:Ccontours9a}
}
\hspace{.5cm}
\subfigure[{\footnotesize \  }]{\includegraphics[width=0.48\textwidth]{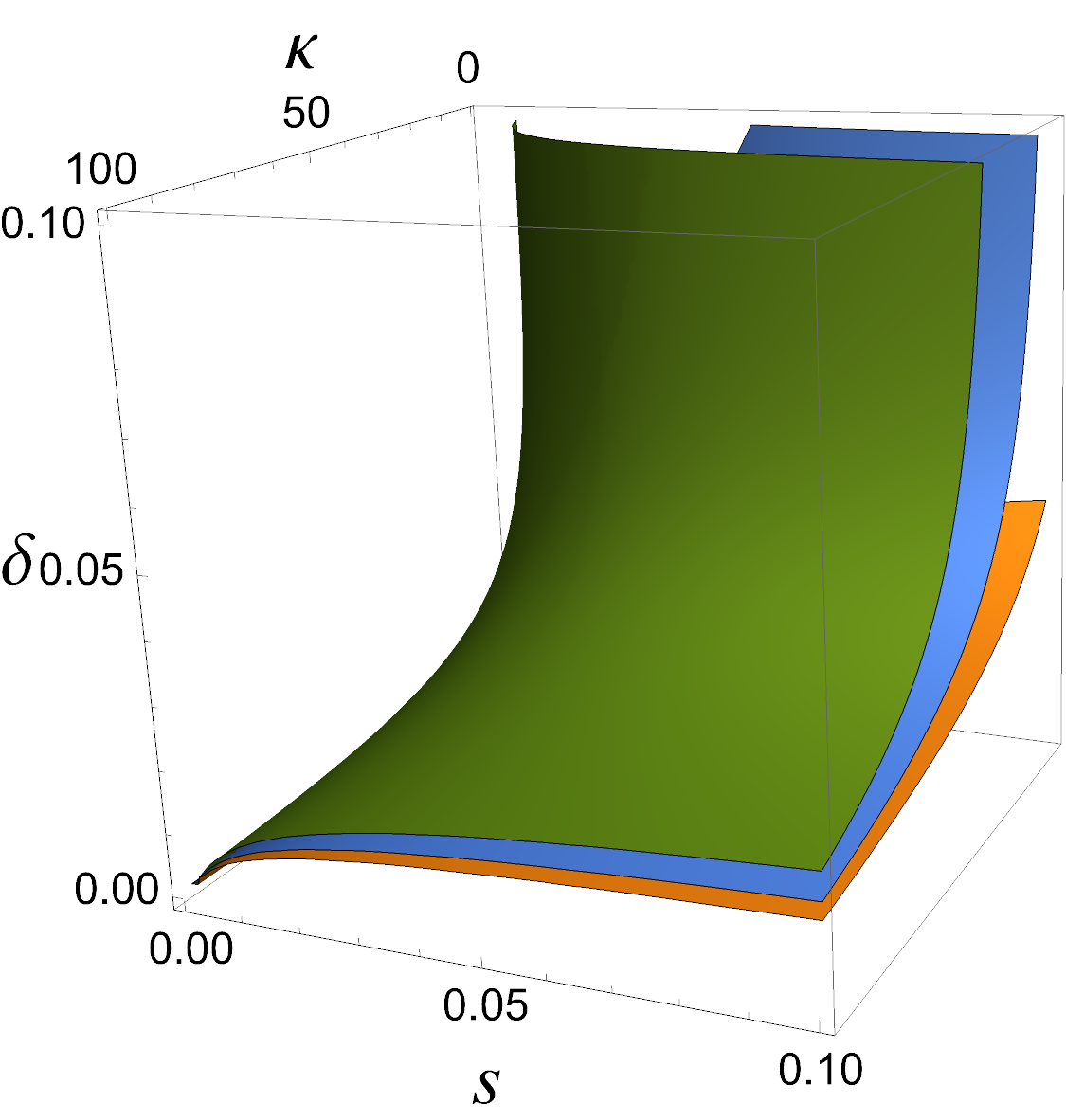}
\label{fig:Ccontours9b}
}
\caption[sm]
{(a) View of the level sets of the  Casimir $C$ for $\al=\pi/9$. (b)    Same as (a) but rotated.
}
\label{fig:Ccontours9}
\end{figure}

\begin{figure}[htb]
\centering
\subfigure[{\footnotesize \  }]{\includegraphics[width=0.45\textwidth]{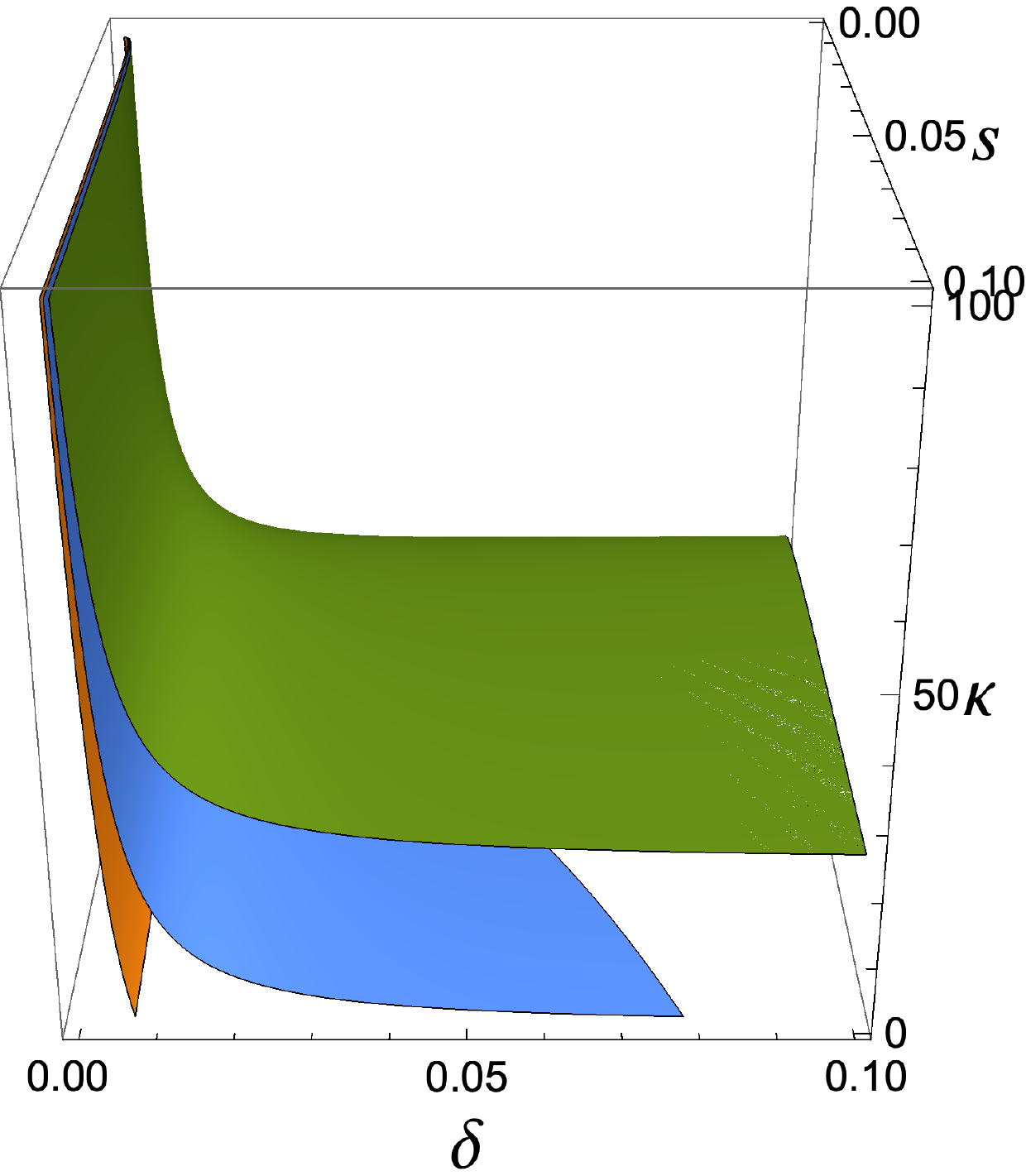}
\label{fig:Ccontours90a}
}
\hspace{.5cm}
\subfigure[{\footnotesize \  }]{\includegraphics[width=0.46\textwidth]{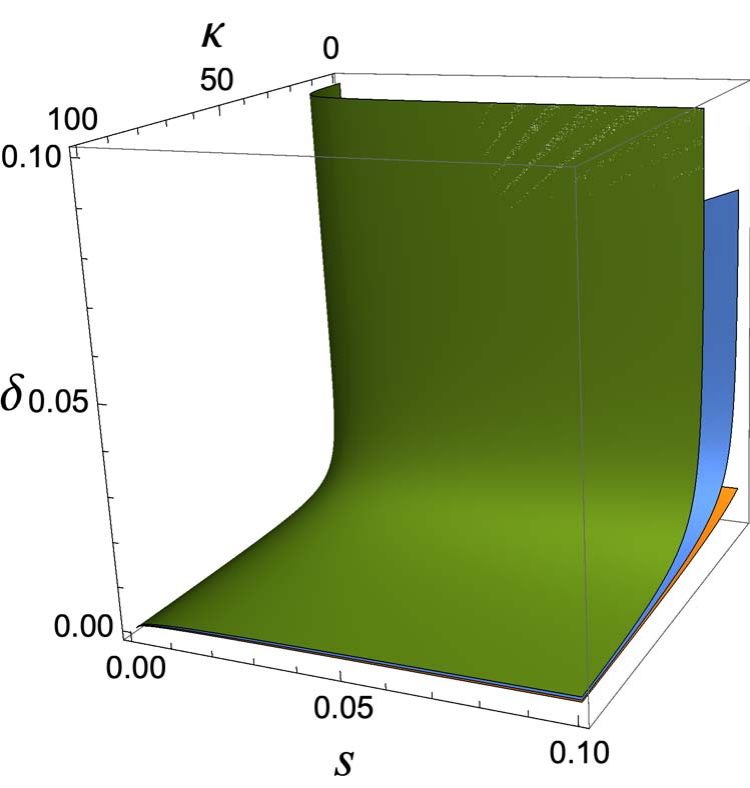}
\label{fig:Ccontours90b}
}
\caption[sm]
{(a) View of the level sets of the  Casimir $C$ for $\al=\pi/90$. (b)    Same as (a) but rotated.
}
\label{fig:Ccontours90}
\end{figure}

In Figs.~\ref{fig:Ccontours4}, 
   \ref{fig:Ccontours9},   and \ref{fig:Ccontours90} 
we show two views of contour plots of the Casimir of \eqref{Cfinal} for three values of the  vortex ring tilt  angle $\alpha\in\{\pi/4, \pi/6,\pi/90\}$. 
Observe that these surfaces are again sheet like but with more interesting structure, no longer having the $\ka$ independence of the $H$-surfaces.  

 The case where $\al=\pi/90$ corresponds to nearly parallel vortices.  For this case  $c_1\approx 5\times 10^{-5}<<1$ and \eqref{dotka0} implies $\ka$ is nearly constant.  For this case the trajectory  lingers in the flat regions of Fig.~\ref{fig:Ccontours90} and the dynamics is approximately governed  by 
\bal
   \dot{\de}&=-\frac{c_2}{2}\,   \frac{\mathring{\ka} \de }{s}
  \label{dotde00}\\
 \dot{s}&=- c_2\, \mathring{\ka}  \left[ \ln\left(\frac{s}{\de}\right) +\be_1\right]\,,
   \label{dots00}
\eal
with $c_2\approx 0.0796$.  Because $H$ is independent of $\ka$, it is not a surprise that \eqref{H}  is still conserved by \eqref{dotde00}  and \eqref{dots00}.  This leads to the quadrature discussed in Appendix \ref{sec:kac}, where the `$\ka$-clock' is proportional to ordinary time. As $s$ gets small,   $\ka$  becomes activated,  demonstrating the importance of the  local induction velocity 
for developing curvature. 
 
\begin{figure}[htb]
\centering
\subfigure[{\footnotesize \  }]{\includegraphics[width=0.47\textwidth]{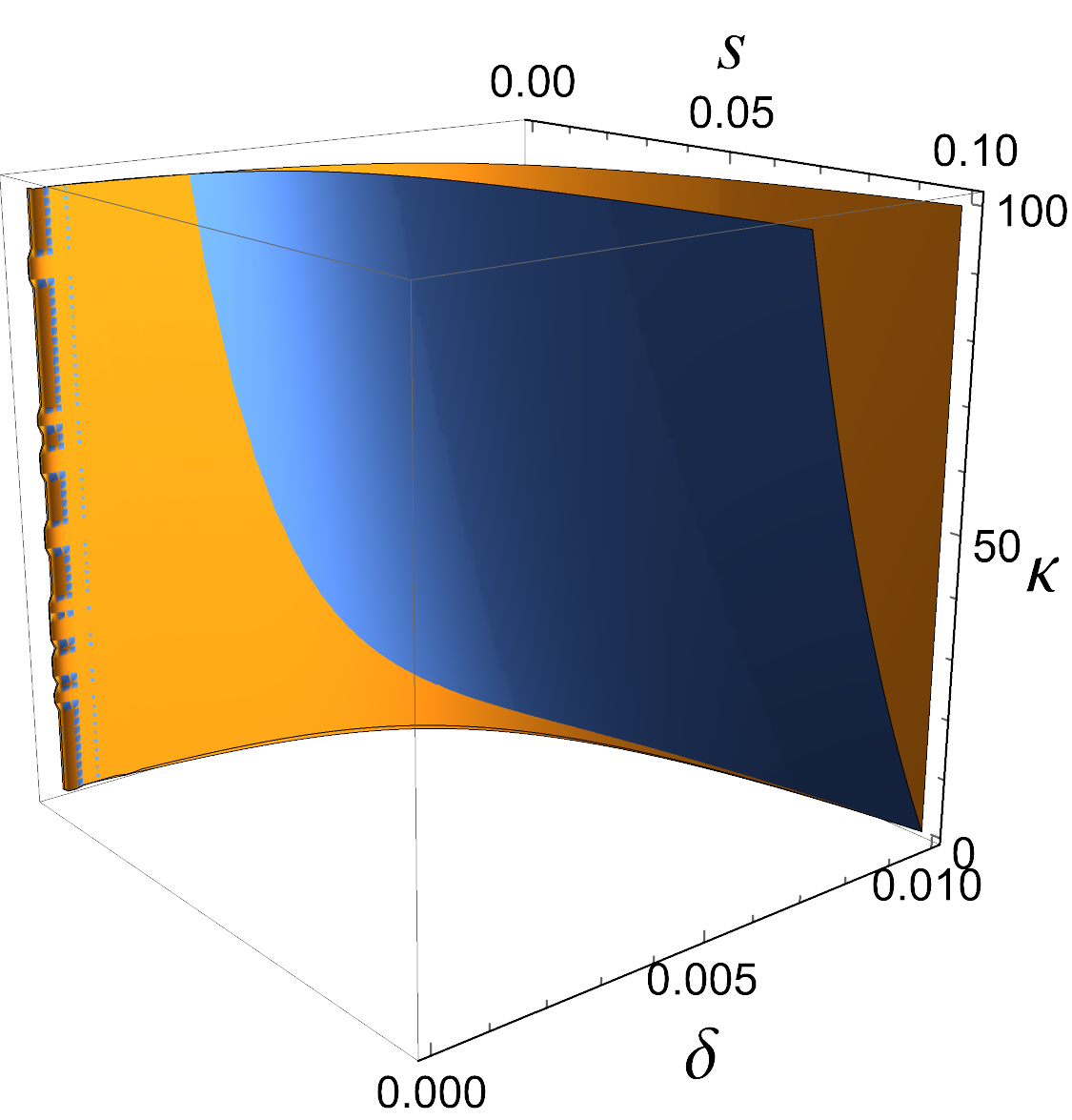}
\label{fig:SIC_100}
}
\hspace{.5cm}
\subfigure[{\footnotesize \  }]{\includegraphics[width=0.47\textwidth]{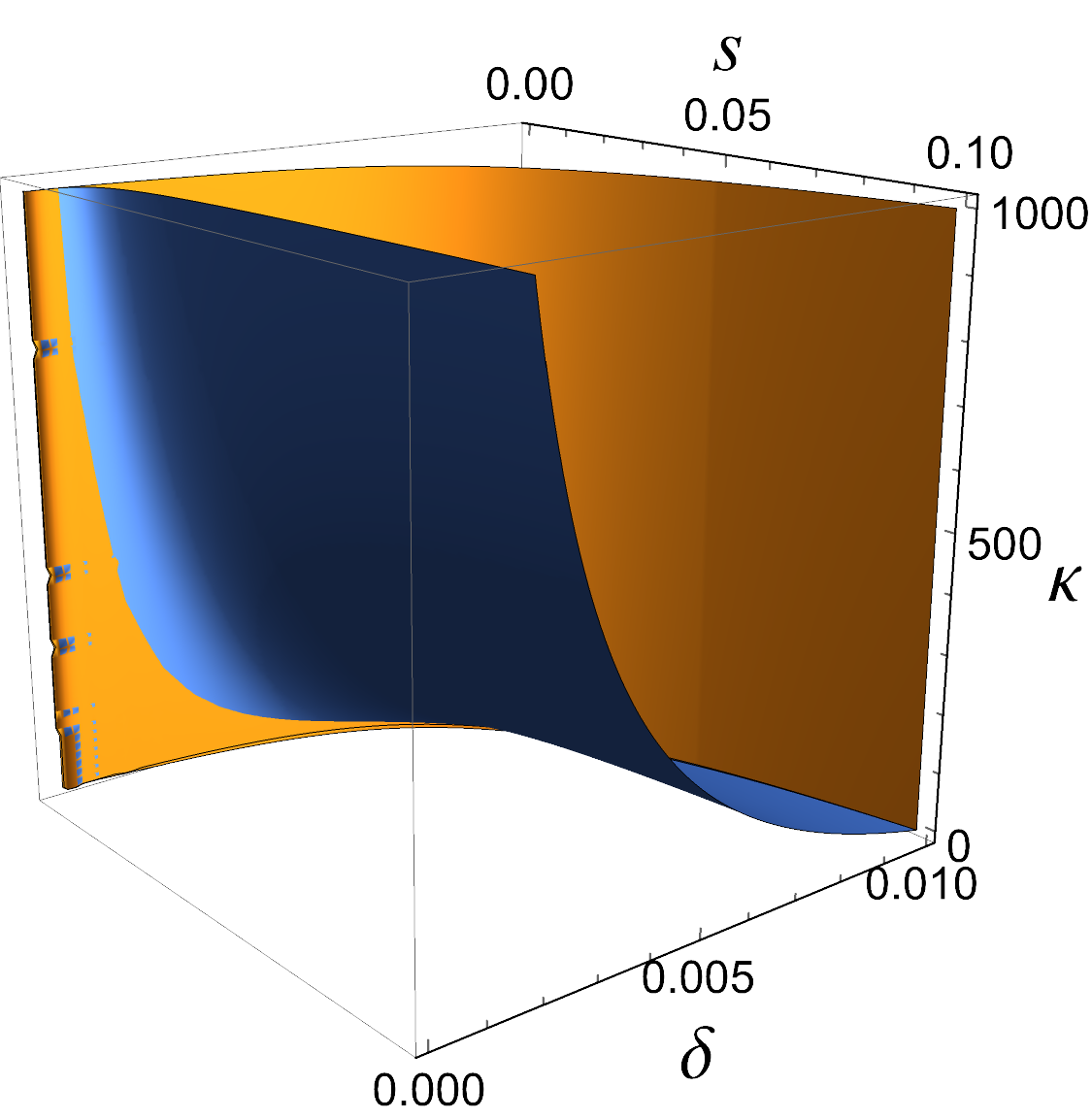}
\label{fig:SIC_1000}
}
\caption[sm]
{Plot of the intersection of the level set $\mathring{H}= 22442.9$ (orange) with that of  $\mathring{C}=-6.40628$ (blue) for $\al=\pi/9$.  These values correspond to the initial conditions of \eqref{SIC}.  In  (a) $\ka$ ranges to 100, while in (b) we zoom out to $\ka$ ranging to 1000  to give a more global perspective on the shape of the surfaces. }
\label{fig:HC9_SIC}
\end{figure}

Now consider the roles played by  the $H$ and $C$ surfaces. As a specific example, consider the case with the initial conditions
\bq
\mathring{\de}=0.01 <  \mathring{s}=0.10 <  \mathring{\ka}=1\,.
\label{SIC}
\eq
With the values of \eqref{SIC} and a choice for the vortex ring tilt angle $\al$,    the solution lies on the intersection of the level sets with 
\bq
\mathring{H}= 22442.9 \andqq \mathring{C}=-6.40628\,,\qquad \mathrm{for}\quad \al=\pi/9\,, 
\eq
where again we use $\be_1=0.4417$.  Thus, these  initial conditions and corresponding initial values of 
$\mathring{H}$ and $\mathring{C}$ are  consistent wtih  \eqref{ineq}.   This particular intersection is displayed in Fig.~\ref{fig:HC9_SIC}.  
Observe how the curve of intersection approaches increasingly  large values of $\ka$ as both $\de$ and $s$ approach zero. 

\begin{figure}[htb]
\centering
\subfigure[{\footnotesize \  }]{\includegraphics[width=0.47\textwidth]{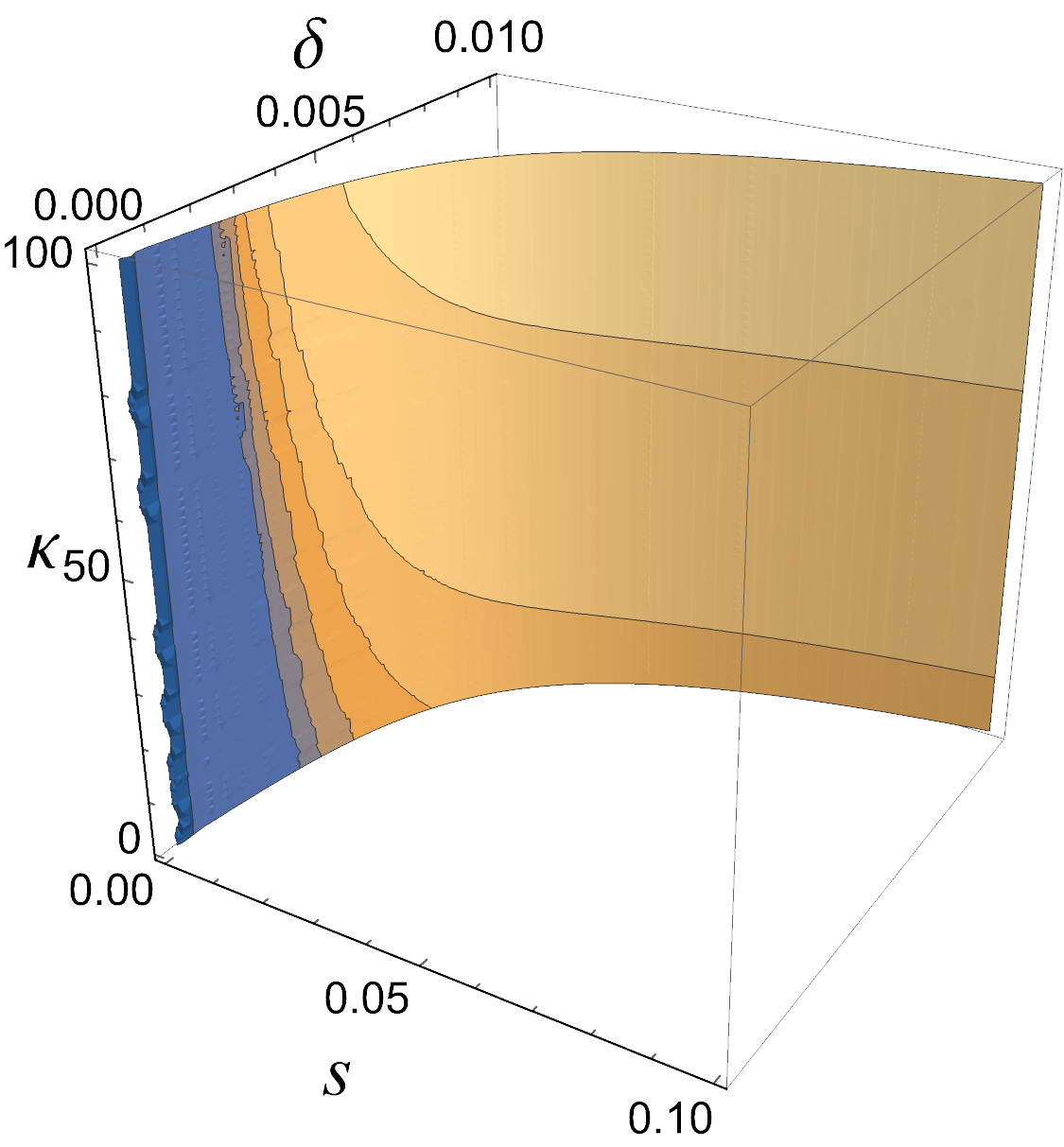}
\label{fig:C0H}
}
\hspace{.5cm}
\subfigure[{\footnotesize \  }]{\includegraphics[width=0.47\textwidth]{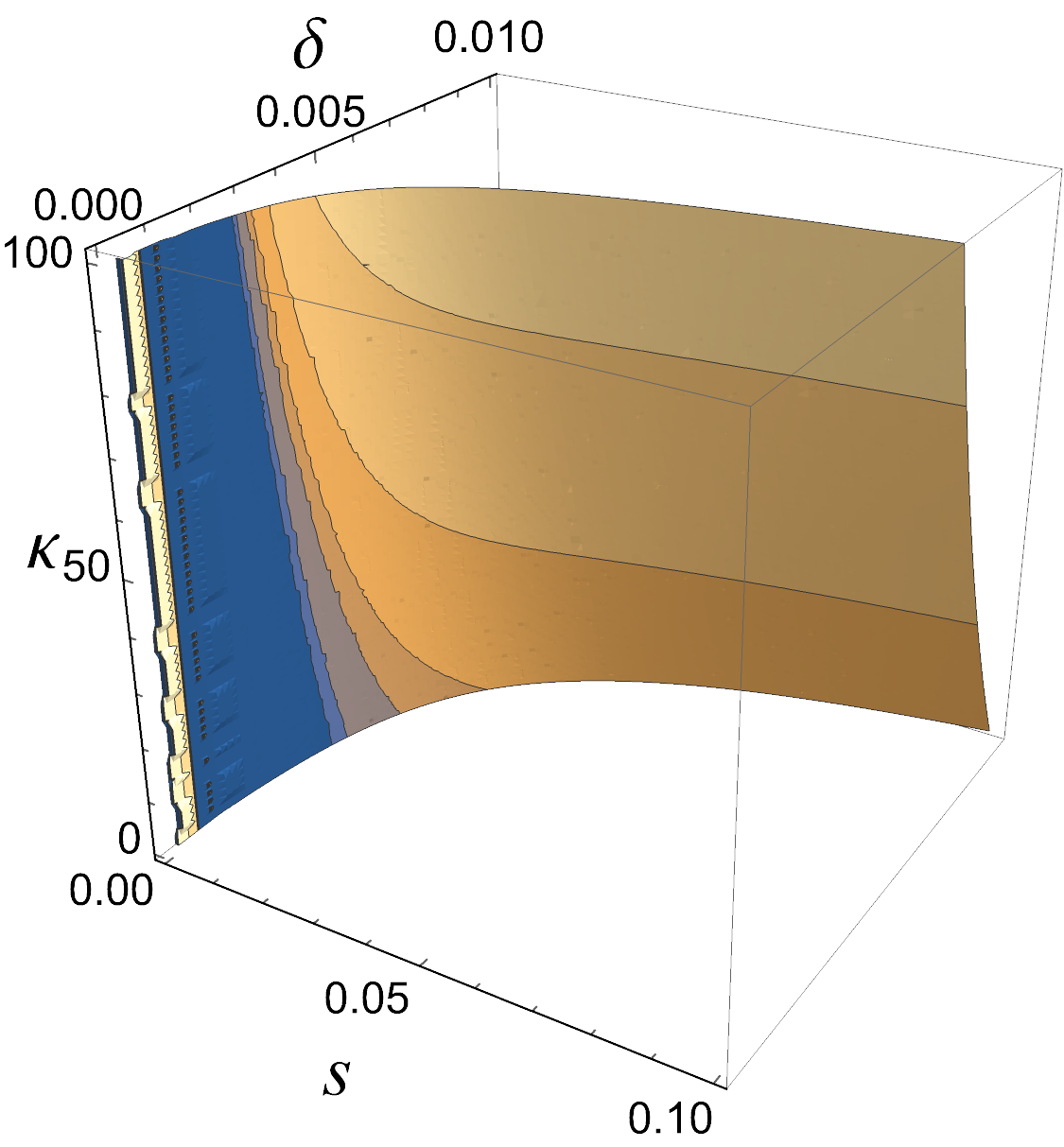}
\label{fig:H0C}
}
\caption[sm]
{Intersections of the level sets of $H$ and $C$  for the initial conditions of \ref{SIC} with $\al=\pi/9$.  In (a) we set 
$\mathring{C}= -6.40628$ and show contours for $\mathring{H}\in\{5,000; 15,000; 20,100; 30,000;45,000\}$, while in (b) we set  $\mathring{H}=22442.9$ and show contours for $\mathring{C}\in\{-10,-9,-8,-7,-6,-4\}$.
}
\label{fig:HCrange}
\end{figure}

In Fig.~\ref{fig:HCrange} we plot multiple intersections of the $H$ and $C$ contours. 
From this figure we see how $\ka$ diverges for a variety of initial conditions,  behavior that is in fact generic.

\begin{figure}[htb]
\includegraphics[scale=.4]{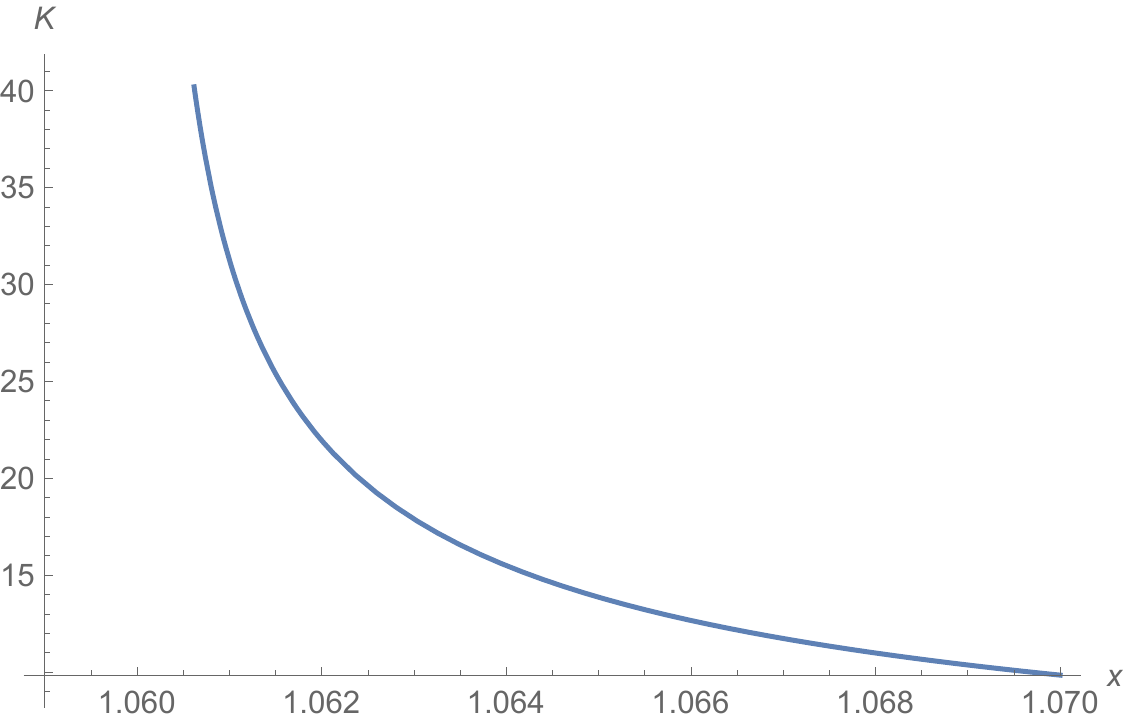}
\caption[sm]{Plots of  the function $K$ of \eqref{Keq} vs.\ $x$, depicting the singularity that occurs for $\ka$ at $x^* \approx 1.0600$.}
\label{fig:K}
\end{figure}

Yet another picture of the singularity emerges if we eliminate $\de$ between \eqref{H} and \eqref{Cfinal}, giving the expression
\bal
\frac{c_2\ka}{2C}  &= 1- \frac{\sqrt{H}}{C} \frac{\la(x)}{\sqrt{\La (x)-1/2}}
\ncr
&=  1+  \frac{c_1\sqrt{H}/{C}} {\sqrt{\La (x)-1/2}}\left(
 \sqrt{\pi} \, e^{\be_1-1/2} \, {\sqrt{\La -1/2\,}}\,  \mathrm{erf}\left(\sqrt{\La -1/2}\right)   
 + \frac{1}{x}
 \right)
 \ncr
 &=:1+ \frac{c_1\sqrt{H}}{C}\, K[x]\,.
 \label{Keq}
\eal
This formula is well defined for $H<0$ because  $\de= \sqrt{H/(\La -1/2)}>0$.   Figure \ref{fig:K} shows the function $K$ with the singularity occurring at $x^* \approx 1.0600$.

The reader may wonder what would happen to the Casimir  if we retained  the term of \eqref{A1} with  the integration constant $c_0$.  It turns out that this  merely adds a term proportional to $\sqrt{H}$ to the Casimir and thus just shifts the value of $C$,  and thus has no consequence other than changing the numerical value of $C$ for the same plots.

\subsection{Reduction to quadrature}

Let us now consider exact integration of the system.  Suppose $(\mathring{\de},\mathring{s}, \mathring{\ka})$ represent an arbitrary initial condition at time $\mathring{t}$ with corresponding initial values of our two invariants $\mathring{H} = H(\mathring{\de},\mathring{s}, \mathring{\ka})$ and $\mathring{C} = C(\mathring{\de},\mathring{s}, \mathring{\ka})$.  Because $H =\mathring{H}$ for all time,  \eqref{H} can be used in \eqref{Cfinal}  to obtain 
\bq
\mathring{C}=\frac{c_2}{2}\ka - c_1 e^{\be_1-1/2} \sqrt{\pi\mathring{H}} \, \,
  \mathrm{erf}\left(\de\sqrt{\mathring{H}}\right)    - \frac{c_1}{s} \,,
  \label{C1}
\eq
where recall  this formula is well-defined when $\mathring{H}<0$ because it  analytically continues to the expression of \eqref{Af2}.   Next,  using \eqref{sde} to eliminate $s$  we  obtain
\bq
\mathring{C}=\frac{c_2}{2}\ka - c_1 e^{\be_1-1/2} \sqrt{\pi\mathring{H}} \, \,
  \mathrm{erf}\left(\de\sqrt{\mathring{H}}\right)    - \frac{c_1}{\de} \,  e^{-\de^2\mathring{H} +\be_1 - 1/2}\,.
  \label{intermed}
\eq
Solving \eqref{intermed} for $\ka$ is immediate
\bq
\frac{c_2}{2}\ka= \mathring{C}  + c_1 e^{\be_1-1/2} \sqrt{\pi\mathring{H}} \, \,
  \mathrm{erf}\left(\de\sqrt{\mathring{H}}\right)   +  \frac{c_1}{\de} \,  e^{-\de^2\mathring{H} +\be_1 - 1/2}\,.
  \label{kade}
\eq
With  \eqref{sde} and \eqref{kade}, we obtain from   \eqref{dotde}, the following:
\bal
\dot{\de}&=-\frac{c_2}{2}\frac{\ka \de }{s} =-
\frac{c_2}{2}\ka \, e^{-\de^2\mathring{H} +\be_1 -1/2}
\ncr
&=- \left( \mathring{C}  + c_1 e^{\be_1-1/2} \sqrt{\pi\mathring{H}} \, \,
  \mathrm{erf}\left(\de\sqrt{\mathring{H}}\right)   +  \frac{c_1}{\de} \,  e^{-\de^2\mathring{H} +\be_1 - 1/2}
\right) \, e^{-\de^2\mathring{H} +\be_1 -1/2}\,,
\label{ddede}
\eal
which leads immediately to  the quadrature
\bq
\mathring{t} - t= e^{-2 \be_1 +1}  \,\,\int_{\mathring{\de}}^\de  
 \frac{ e^{\de'^2\mathring{H}}\, \,\de'\,   d \de'}
{\de'\,e^{-\be_1 +1/2}\mathring{C}   + c_1  \sqrt{\pi\mathring{H}}\de' \, \,
  \mathrm{erf}\left(\de'\sqrt{\mathring{H}}\right)   +    c_1 e^{-\de'^2\mathring{H}} 
} \,.
\label{dequad}
\eq
Although unwieldy, integration of \eqref{dequad} gives $\de$ as a function of time, and  via  \eqref{sde} and \eqref{kade}  we obtain $s$ and $\ka$ as functions of time, with the latter diverging. 

 From \eqref{dequad} we see that if there is a finite-time singularity where $\de\rightarrow 0$, then it occurs at a time $t_\infty$ in accordance with the following formula:
\bq
t_\infty=\mathring{t} +  e^{-2 \be_1 +1}  \,\,\int^{\mathring{\de}}_0 
 \frac{ e^{\de'^2\mathring{H}}\, \,\de'\,   d \de'}
{\de'\,e^{-\be_1 +1/2}\mathring{C}  +  c_1  \sqrt{\pi\mathring{H}}\de' \, \,
  \mathrm{erf}\left(\de'\sqrt{\mathring{H}}\right)   +   c_1  e^{-\de'^2\mathring{H}} 
} \,.
\label{tinfty}
\eq
For $\mathring{H}\neq 0$,  we see from \eqref{H} that at  a singularity where $\de\rightarrow 0$, we must have $\La -1/2\rightarrow 0$.  As noted in \eqref{critH}, 
 the smallest  initial value of $H$ that satisfies   \eqref{ineq} is $\mathring{H}^*= (\be_1-1/2)/\mathring{\de}^2$, which is negative;
 if  $\mathring{H}> \mathring{H}^*$, then 
\eqref{ineq} is satisfied initially.  If $\de$ decreases  and it initially satisfies   $s>\de$,  then it will satisfy it throughout its evolution  because  $\de=s\, e^{-\de^2\mathring{H} +\be_1 -1/2}$.

\subsection{Exact  solutions with Leray scaling}
\label{ssec:exleray}

Evidently, the integral of \eqref{dequad} is dramatically simplified with the choice  $\mathring{H}=0$.  
With this choice,  \eqref{sde} implies  
\bq
s=\de e^{-\be_1 + 1/2} 
\label{salde}
\eq
and equations \eqref{dotde} and \eqref{dots} become identical upon setting  
\[
\La  =\ln(s/\de) +\be_1=1/2\,, 
\]
a choice  consistent with  the inequality of \eqref{ineq}.  
Given that for gaussian core profiles $\be_1=0.4417$  (see  \cite{MK19}), we have 
\bq
  e^{-\be_1 + 1/2}  = 1.0600\quad\Rightarrow \quad s>\de \,.
\eq
Thus inequality \eqref{ineq} is satisfied, although only barely.   Proceeding, the  MK system reduces for $\mathring{H}=0$  to 
 \bq
 \dot{s}=- \frac{c_2}{2}\,  \ka  
\andqq
 \dot{\ka}=c_1\,  \frac{\ka}{s^2}\,,
  \label{nudotka}
\eq
with the Casimir becoming 
\bq
C= \frac{c_2}{2} \ka-\frac{c_1}{s}\,,
\label{CH0}
\eq
which follows from \eqref{C1}. 

Before considering arbitrary initial $\mathring{C}$, we consider the easily tractable case where both  $\mathring{H}=0$ and $\mathring{C}=0$.  This   implies
\bq
s\ka= 2c_1/c_2 = 2\sin\al\,.
\label{ska}
\eq
Therefore if we choose 
\bq
2\sin\al <1\qquad \mathrm{or}\qquad \al<\frac{\pi}{6}\,, 
\eq
then the other part of the  inequality   \eqref{ineq} is satisfied, viz.\  $s<1/\ka$.
 Note, the value of $\al=\pi/4$ chosen in \cite{MK19,MK19b} does not satisfy this inequality. 
Using \eqref{ska} in the  $\dot{\ka}$ equation of  \eqref{nudotka} gives 
\bq 
\dot{\ka}=\frac{c_2^2}{4 c_1}  \ka^3 \,,
\eq
which is easily integrated to obtain
\bq
- \frac{1}{2\ka^2} +  \frac{1}{2\mathring{\ka}^2}  = \frac{c_2^2}{4 c_1}(t-\mathring{t})
\eq
and  the exact solution 
\bq
\ka^{-1} = \sqrt{\frac{c_2^2}{2c_1}(t_\infty-t)}= \sqrt{\frac{\mathrm{cot}\al}{8\pi}(t_\infty-t)}\,,
\label{lerayE}
\eq
where 
\bq
t_{\infty}= 
\mathring{t} + \frac{2c_1}{c_2^2\mathring{\ka}^2} 
=\mathring{t} + \frac{8\pi \tan\al }{\mathring{\ka}^2}
\,.
\label{tinfytC0}
\eq 
As expected, smaller values of $\ka$ take longer to diverge. 

The rest of the solution is obtained from \eqref{ska} and \eqref{salde}, i.e., 
\bal
s&=2 \ka^{-1} \, \sin\al= 2\sin\al \sqrt{\frac{\mathrm{cot}\al}{8\pi}(t_\infty-t)}
\ncr
&= \sqrt{\frac{\sin(2\al)}{4\pi}(t_\infty-t)}=\de e^{1/2-\be_1} \,.
\label{sdeleray}
\eal
 It is a simple matter to insert the solutions of \eqref{lerayE} and \eqref{sdeleray} into \eqref{dotde}, \eqref{dots}, and  \eqref{dotka} to verify directly that they are indeed an exact solution, one that satisfies the inequalities of \eqref{ineq}.

Now consider the more general case $\mathring{C}\neq 0$.   Solving \eqref{CH0} for $s$ and inserting into \eqref{nudotka} gives 
\bq 
\dot{\ka}=\frac{\ka}{c_1}  \left(\frac{c_2}{2}\ka - \mathring{C} \right)^2
= \frac{c_2^2}{4c_1}\ka \left(\ka - 2\mathring{C}/c_2 \right)^2\,, 
\eq
which is easily integrated to obtain the exact solution, 
\bq
  \ln\Big(1-\frac{2\mathring{C}/c_2}{\ka}\Big) +  \frac{2 \mathring{C}/c_2 }{\ka\, (1 -   \frac{2 \mathring{C}/c_2}{\ka} )}
  - \ln\Big(1-\frac{2\mathring{C}/c_2}{\mathring{\ka}}\Big) - \frac{2 \mathring{C}/c_2 }{\mathring{\ka}\, (1 -   \frac{2 \mathring{C}/c_2}{\mathring{\ka}})}
= - \frac{\mathring{C}^2}{c_1} \,(t - \mathring{t})\,.
\label{kasol}
\eq
 Because the model only makes sense if $\de, s$, and $\ka$ are all greater than or equal to zero, we must have $c_1/s=\ka{c_2}/{2} -C\geq 0$ or $1-2 {C}/( {\ka} c_2)\geq 0$, which  in fact according to \eqref{Cineq}  is true for all allowable values of $\mathring{H}$.
 
\begin{figure}[htb]
\centering
\subfigure[{\footnotesize \  }]{\includegraphics[width=0.45\textwidth]{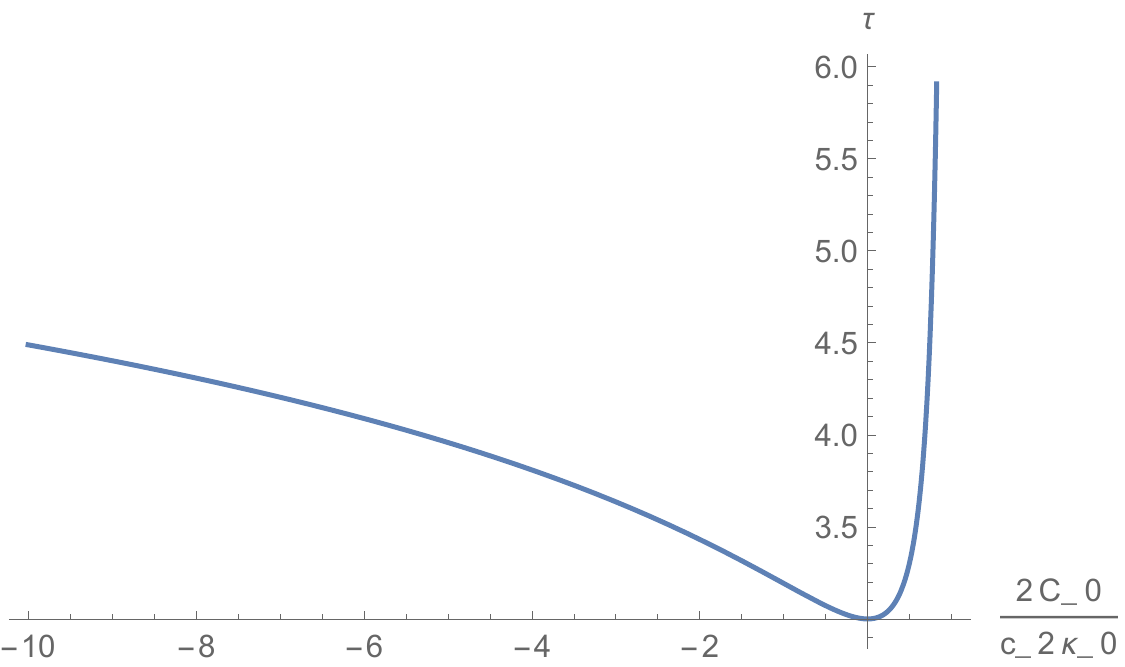}
\label{fig:blowupH0}
}
\hspace{.5cm}
\subfigure[{\footnotesize \  }]{\includegraphics[width=0.45\textwidth]{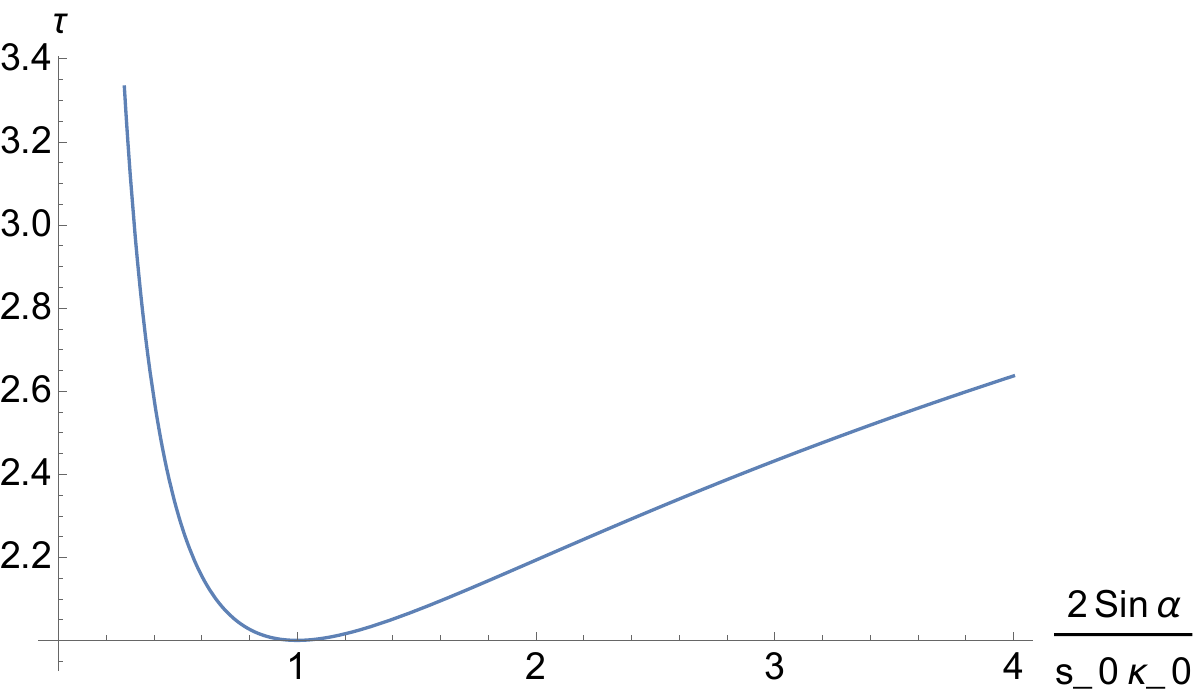}
\label{fig:blowup_again} 
}
\caption[sm]
{Plots of the blowup times for $ \mathring{H}=0$, $ \mathring{C}\neq 0$, (a) according to Eq.~\eqref{blowupH0} where   $\tau= (t_\infty- \mathring{t})\mathring{C}^2/c_1$  vs.\  $2 \mathring{C}/(c_2\mathring{\ka})$ and  (b)  according to Eq.~\eqref{blowupH0} where    $\tau= (t_\infty- \mathring{t})\mathring{C}^2/c_1$ vs.\  $X={2\sin\al}/({\mathring{\ka}\mathring{s}})$.
}
\label{F:blowups}
\end{figure}

Assuming the physical initial conditions  satisfy $\mathring{\ka}>  2   \mathring{C}/c_2$, we see the $\dot\ka>0$, so $\ka$ continues to grow,    with divergence occurring at the finite time 
\bq
t_{\infty}=\mathring{t} + \frac{c_1}{\mathring{C}^2}\left( 
   \ln\Big(1-\frac{2\mathring{C}/c_2}{\mathring{\ka}}\Big) +  \frac{2 \mathring{C}/c_2 }{\mathring{\ka}\, (1 -   \frac{2 \mathring{C}/c_2}{\mathring{\ka}} )}
\right)\,,
\label{blowupH0}
\eq
an approximation to \eqref{tinfty} but a generalization to \eqref{tinfytC0}.  This is depicted in Fig.~\ref{fig:blowupH0},

The quantity $t_{\infty}$ can also be written as a function of $\ka s$, which makes it convenient for assessing initial conditions compatible with \eqref{ineq}.  Using  \eqref{CH0} we obtain
\bq
t_{\infty}=  \mathring{t} + \frac{c_1}{\mathring{C}^2} \left(\ln(X) +\frac{1}{X} - 1 \right)\,,
\eq
where
\bq
X= \frac{2c_1}{c_2\mathring{\ka}\mathring{s}} =\frac{2\sin\al}{\mathring{\ka}\mathring{s}} \,.
\eq
Observe from Fig.~\ref{fig:blowupH0}  there is blow up for all values of $\mathring{C}$, but only values consistent with 
$\mathring{s}<1/\mathring{\ka}$ are acceptable. Thus, 
\bq
\quad \frac{1}{\mathring{s}\mathring{\ka}} >1\quad\Rightarrow \quad
 \frac{2\sin\al}{\mathring{s}\mathring{\ka}} >2\sin\al\,.
\eq
Therefore after choosing  $2\sin\al<1$, only blowup times to the right of this value in Fig.~\ref{fig:blowup_again} are consistent with \eqref{ineq}.

As for the case  $\mathring{C}=0$, given the solution of \eqref{kasol} for $\ka$,  we immediately obtain the solutions for $s$ and $\de$ 
 from \eqref{CH0} and \eqref{salde}, respectively, 
\bq
s= \frac{2c_1/c_2}{\ka - 2\mathring{C}/c_2}= \de\, e^{-\be_1 + 1/2} \,.
\label{sdeexact}
\eq
 
At late times, expansion of  \eqref{kasol}  again gives 
\bq
\ka^{-1} \sim \sqrt{\frac{c_2^2}{2c_1}(t_\infty-t)}\,,
\label{lerayE2}
\eq
in agreement with the  Leray scaling  suggested in \cite{MK19} and proven above in \eqref{lerayE} and \eqref{sdeleray}.   Note, in \eqref{lerayE},   the constant $\mathring{C}$ only appears via  $t_\infty$.  The zero energy case is special in that it is tractable, but the Leray scaling is ubiquitous.    Inserting \eqref{lerayE} into \eqref{sdeexact} we obtain for late times
\bq
s =  \de\, e^{-\be_1 + 1/2}  \sim  \sqrt{2c_1(t_\infty -t)}
\label{sleray}
\eq
and so  $\ka s\sim 2c_2/c_1$ as for the case with $\mathring{C}=0$.

 \subsection{General analysis}
 \label{ssec:genanal}
 
Let us now return to  \eqref{dequad} and consider the general solution.  Because  $\mathring{C}$ is arbitrary, it follows immediately from \eqref{ddede}  that there is a family of equilibrium  points of $\de$ when the righthand side vanishes.   However, by  \eqref{kade},   all of these correspond to $\ka=0$ (see  Appendix \ref{sec:NHS} for an analysis).  Also note, for $\ka>0$, $\dot\de <0$ and so even when $\mathring{H}\neq 0$,  we  expect a solution where $\de$ approaches  zero.  If indeed $\de\rightarrow 0$, then \eqref{sde} implies
\bq
s\approx \de e^{ -\be_1 +1/2} +\calo(\de^3)
\eq
for all $\mathring{H}$; i.e., as for $\mathring{H}=0$, $\de$ must go to zero with $s$ along the line
\bq
s= \de e^{ -\be_1 +1/2} \,, 
\eq
while from \eqref{kade} $\ka$ must diverge as 
\bq
\frac{c_2}{2}\ka  = \mathring{C}   +  \frac{c_1}{\de} \,  e^{\be_1 - 1/2}\,.
  \label{kade2}
\eq
Note,   the sign of the righthand side of \eqref{ddede} is definite provided $\mathring{C}>0$ and $\de>0$.

Let us explore further the behavior for small $\de$.  To this end we use 
\[
\mathrm{erf}(\sqrt{y})
= \frac{2}{\sqrt{\pi}} \sum_{n=0}^\infty \frac{(-1)^n \, y^{n+1/2}}{n!\,(2n+1)}\,, 
\]
for small $y\in\R_+$,   $\sqrt{y}\, \mathrm{erf}(\sqrt{y})\in C^\om$,  and the  identities are
\bq
\frac{d}{dx}  \mathrm{erf}(x)=  \frac{2}{\sqrt{\pi}}\,e^{-x^2}
\andq
\mathrm{erf}(x)=\frac{d}{dx} \left(x\, \mathrm{erf}(x) + \frac{e^{-x^2}}{\sqrt{\pi}}\right)\,.
\label{useerfid}
\eq
Upon multiplying \eqref{ddede} by $\de$ and defining $u=\de^2/2$, it  takes the form
\bq
\dot{u}= u^{\frac12}g(u) - f(u)
\eq
where $f,g\in C^\om(\R)$  are defined by
\bal
f(u)&= e^{-2\mathring{H}u+2\be_1 - 1}c_1\left( \sqrt{2\pi\mathring{H} u\,} \,
  \mathrm{erf}\left(\sqrt{2\mathring{H}u\,}\right)   +    e^{-2\mathring{H}u}
\right)
\label{fu}
\\
g(u)&=- e^{-2\mathring{H}u+\be_1 - 1/2} \sqrt{2} \, \mathring{C}\,,
\label{gu}
\eal
and  satisfy
\bq
f(0)= c_1e^{2\be_1-1}\qquad\mathrm{and}\qquad 
g(0)=-\sqrt{2} \, \mathring{C}\, e^{\be_1 -1/2}\,.
\eq
Using \eqref{useerfid}  we see that the function  $f(u)$ is positive and monotonically decreasing, while the function $g(u)$ has  a sign determined by the sign of $\mathring{C}$,  and is at least  monotonic for small argument.

\begin{figure}[htb]
\includegraphics[scale=.6]{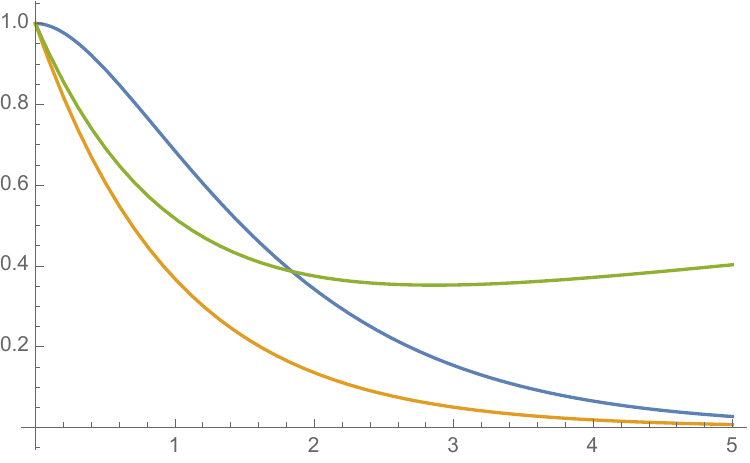}
\caption[sm]{Plots of  $\bar{f}=  e^{-x} (\sqrt{\pi x}\, \mathrm{erf}( \sqrt{x}) +e^{-x})$ (blue), $\bar{g}= 0.10 \times e^{-x}$ (orange),  and  $f/g=\sqrt{\pi x}\, \mathrm{erf}(\sqrt{x}) +e^{-x}$ (green) vs.\  $x=2\mathring{H}u$. Note,  $\sqrt{u_\infty} \sim f(u_\infty)/g(u_\infty)$ are the equilibria points corresponding to $\ka=0$.}
\label{f&g}
\end{figure}

The functions $f$ and $g$ are depicted in Fig.~\ref{f&g}. 

Thus, near $u=0$ the MK system behaves as 
\bq
\dot{u}= u^{\frac12}g(0) - f(0)
\label{dotsqu}
\eq
Because $u^{\frac12}g(0) - f(0)$ is not Lipschitz on $u\in[0,\ep]$ with $\ep>0$, we do not have the usual ODE existence theorem to rely on (see e.g.\  \cite{c&l}).  If $u(0)=\mathring{u}>0$ is small, then $\dot{u}(0)= \mathring{u}^{\frac12}g(0) - f(0)<0$, because $c_1>0$, and  $u$ should decrease. 
However, the Lipschitz condition for uniqueness is  a sufficient but not a necessary condition.  Thus, further analysis is necessary, but indeed  
\eqref{dotsqu} does have a unique solution. To see this let  $u_1$ and $u_2$ be  two solutions that coincide at some time.   A measure of their  difference
\bq
D = \left(\sqrt{u_1(t)} -\sqrt{u_2(t)}\right)^2
\eq
satisfies $\dot D=0$ for all time,  so they must coincide. 

\bigskip

From \eqref{dotsqu} we see there is a family of equilibrium points given by 
\bq
 \sqrt{u_0} = \frac{f(0)}{g(0)}=  \frac{c_1e^{\be_1-1/2}}{ \sqrt{2}\, \mathring{C} }
 \eq
which only exists for $\mathring{C}>0$.  However, these are merely the $\ka=0$ solutions mentioned above.  
In Appendix \ref{sec:NHS} it is seen that linearization about any of these equilibrium solutions  yields a spectrum with   two zero eigenvalues and one unstable (positive) eigenvalue.   This implies $\ka$ will grow so as to decrease the radius of curvature.

In any event,  it is a simple matter to integrate \eqref{dotsqu}, yielding
\bq
 \frac{2}{g(0)}\left[
 u^{\frac12} + \frac{f(0)}{g(0)} \ln\left(1-  {u^{\frac12} g(0)}/{f(0)}\right) 
 \right]= t + \mathrm{const}\,,
\label{Cgeq0}
\eq
 which, upon expanding,  gives for small $u$ 
\bq
\de\sim   \sqrt{2f(0)({t_\infty-t)}}=e^{\be_1-1/2}\sqrt{2 c_1(t_\infty-t)}\,.
\eq
the  solution with Leray scaling consistent with \eqref{sleray} and \eqref{lerayE}.

\section{Conclusions}
\label{sec:conclu}

We have shown that the system put forth  by  Moffat and Kimura \cite{MK19,MK19b} for the interaction of two tilted vortex rings has solutions with  finite-time singularity.  This was  achieved by finding the noncanonical Hamiltonian structure of the equations, which naturally led to a geometrical depiction and explicit forms for the solutions by making use of the newly discovered Hamiltonian and Casimir invariants.   Exact Leray divergence was demonstrated, within the inequalities proposed in \cite{MK19,MK19b} for the  model.

Several avenues for future work remain: in future  publications we will consider the effect of viscous dissipation, further physical interpretation of the results,  and various bounds and perturbation expansions.  Of particular interest is to derive the Hamiltonian structure that we have obtained for the  Moffat and Kimura model from that of Euler's equation.  This will elucidate how the Hamiltonian and Casimir of the reduced model relate to those of the parent model.  Insights about vortex lines and Casimirs  described in \cite{K&R},  may be of particular help in this regard.

\section*{Acknowledgment}
\noindent   PJM was supported by U.S. Dept.\ of Energy Contract \# DE-FG05-80ET-53088 and a Forschungspreis from the Alexander von Humboldt Foundation.  YK acknowledges support from JSPS KAKENHI grant \#19H00641 and \#16H06339.

\appendix

\section{Non-Hamiltonian spectrum}
\label{sec:NHS}

Usually one studies the equilibrium  points of a dynamical system in order to get a view into the nature of trajectories in phase space.  Thus we study the only equilibrium point  of  \eqref{dotde0}, \eqref{dots0}, and \eqref{dotka0}, viz.\  $\ka=0$, for any values of $\de$ and $s$, which corresponds to the vortex rings having infinite radii.  Interestingly, although our system of equations \eqref{dotde0}, \eqref{dots0}, and \eqref{dotka0}   is a Hamiltonian system, it has an associated singular Poisson tensor at this equilibrium point.  As discussed in \cite{pjmYT17,pjmY20}, such systems may not have the usual Hamiltonian spectra when expanded about an equilibrium state,  i.e.,  the symmetry of growing and decaying eigenvalues having the same magnitude. 

The singularity occurs at $\ka=0$ for any values of $\de$ and $s$.  Such co-dimension one  singularities may   have the peculiar spectra. This kind of singularity follows because the Poisson tensor  $J$, as given by $\bfV$,  vanishes identically for $\ka=0$.  This is seen because the components $V_i$ for $i=1,2,3$ are all proportional to $\ka$. At other points of phase space it has rank 2, while along this line rank zero. 

The only equilibria of our system occur along the line $\ka=0$, while any values of $\de$ and $s$ are allowed.  Thus expanding as 
\bq
s=s_0 +\tilde s\,,\qquad \de = \de_0 + \tilde \de\,,\qquad  \mathrm{and}\qquad \ka=\tilde\ka
\eq
we obtain a simple eigenvalue problem 
\bq
 \begin{bmatrix} 
\dot{\tilde\de}  \\
\dot{\tilde{s}} \\
\dot{\tilde\ka}\\
\end{bmatrix}
= 
 \begin{bmatrix} 
\, 0&\  0\, & a\,\\
\,0&\  0\,&  b\,\\
\,0&\  0\,& L\,\\
\end{bmatrix}
 \begin{bmatrix} 
\tilde\de \\
\tilde{s} \\
\tilde\ka \\
\end{bmatrix}\,,
\label{linear}
\eq
where
\[
a=-\frac{c_2\de_0}{2 s_0} \,, \qquad b = -c_2 (\ln(s_0/\de_0) + \be_1)\,,\qquad  \mathrm{and}\qquad L=\frac{c_1}{s_0^2}\,, 
\]
with the matrix 
\[
\M:=
 \begin{bmatrix} 
\,0&\  0\, & a\,\\
\,0&\  0\,&  b\,\\
\,0&\  0\,& L \,\\
\end{bmatrix}
\]
giving rise to following characteristic polynomial by assuming temporal behavior of $e^{\ga t}$ 
\bq
\ga^2(\ga- c_1/s_0^2)=0\,.
\eq
Thus the spectrum of $M$ is $\{0,0, c_1/s_0^2\}$. 

In the right coordinates $\M$ is the  direct sum of  commutating diagonal (semisimple) and nilpotent pieces.  To this end we change coordinates by replacing $\tilde\de$ and $\tilde s$, while retaining $\tilde \ka$,  as follows;
\[
\bar\de= \tilde\de -\frac{a}{L}\tilde\ka\qquad \mathrm{and}\qquad \bar{s} :=\tilde s - \frac{b}{L} \tilde \ka
\]
in which case $\M$ is replaced by 
\[
\bar{\M}:=
 \begin{bmatrix} 
\,0&\  0\, & 0\,\\
\,0&\  0\,& 0 \,\\
\,0&\  0\,& L\,\\
\end{bmatrix}
\]
and the linear dynamics is trivial.  With $\tilde\ka$ exponentiating away at fixed initial $\bar\de$ and $\bar{s}$.   From this we conclude that the rings will initially exponentially decrease their radii of curvature while  $\de$ and $s$ decrease.

 \section{The Leray Hamiltonian}
 \label{ssec:lerayHam}
 
In light of Sec.~\ref{ssec:genanal}, we observed  that for small $\de$ the system exhibits Leray scaling.   This behavior follows  upon expanding \eqref{dequad} or by approximating  \eqref{dotde0}, \eqref{dots0}\,, and \eqref{dotka0}.    Here we follow the second route, using in the vicinity of the singularity $s\sim e^{1/2-\be_1} \de=: r \de$, to  obtain  
 the following set of equations that describe the dynamics near the singularity:
\bal
\dot{\de}&= -\frac{c_2}{2 r} \ka +\frac{\varep}{2\de}
\label{redde}
\\
\dot{\ka} &= \frac{c_1}{r^2}\frac{\ka}{\de^2} \,.
\label{redka}
\eal
Upon setting $\varep=0$ (future work considers its retention), we expect this reduced system of \eqref{redde}  and \eqref{redka} to be Hamiltonian.  Indeed, it  conserves the quantity
\bq
\mathfrak{c}= \frac{c_2}{2} \ka - \frac{c_1}{r \de}\,,
\label{tC}
\eq
which upon using  \eqref{Af} and \eqref{lam} can be shown to be the Casimir of  \eqref{C} expanded to leading order,  and \eqref{redde} and \eqref{redka}  can be written in the noncanonical Hamiltonian form as 
\bq
\begin{bmatrix} 
\dot{\de}\\
\dot{\ka}
\end{bmatrix}
=
 \begin{bmatrix} 
0&-  {\ka}/{r} \\
 {\ka}/{r}\ & 0 \\
\end{bmatrix}
 \begin{bmatrix} 
 \p \mathfrak{c}/{\p \de} \\
\p \mathfrak{c}/{\p \ka} \\
\end{bmatrix}
=
 \begin{bmatrix} 
0&-  {\ka}/{r} \\
 {\ka}/{r}\ & 0 \\
\end{bmatrix}
 \begin{bmatrix} 
 {c_1}/({r\de^2)} \\
 {c_2}/{2} \\
\end{bmatrix}\,.
\eq
Upon changing variables according to 
\bq
q=r \de\qquad\mathrm{and}\qquad p= -\ln(\ka)
\eq
 the following canonical Hamiltonian form is obtained:
\bq
\dot{q} = \frac{\p \mathfrak{c}}{\p p} = -\frac{c_2}{2}e^{-p}  
\andqq
\dot{p}=- \frac{\p \mathfrak{c}}{\p q} = - \frac{c_1}{q^2} 
\eq
where the Hamiltonian in canonical coordinates is 
\bq
\mathfrak{c}= \frac{c_2}{2}e^{-p} - \frac{c_1}{q}\,. 
\label{natHam}
\eq
Various  canonical coordinate changes  are possible,   but given that there is a simple  quadrature,  they do not  add  insight.  Clearly, upon setting $\mathfrak{c}$ to a constant, $\mathring{\mathfrak{c}}$, and solving for $p$,  as is usual for natural Hamiltonians of the form of \eqref{natHam},  we obtain the quadrature 
\bq
\int \frac{dq}{\mathring{\mathfrak{c}}q + c_1}= -\int dt\,, 
\eq
leading, yet again,  to  the Leray solution 
\bq
q= r \de \sim\sqrt{2c_1(t_\infty-t)\, }\,, 
\eq
 with $\ka$ following  from $\mathfrak{c}=\mathring{\mathfrak{c}}$, 
\bq
\mathring{\mathfrak{c}}+\frac{c_1}{q}\sim \frac{c_1}{q}  =\frac{c_2}{2} e^{-p}= \frac{c_2}{2} \ka\qquad \Rightarrow \qquad
\ka\sim \sqrt{\frac{2c_1}{c_2^2(t_\infty -t)}}\,.
\eq


\section{Using $\ka$ as a clock}
\label{sec:kac}

There are various paths to quadrature.  Here we present one where the  system is  transformed so that $\ka$ measures time. This is done  by dividing \eqref{dotde} and  \eqref{dots} by  \eqref{dotka}, giving 
 \bqy
   \frac{d \de}{d \bar{\ka}}&=&-  s  \de = - x\de^2
  \label{dkade}\\
 \frac{d s}{d \bar{\ka}}&=&-  s^2  \left[ \ln\left(\frac{s}{\de}\right) +\be_1\right]
 =-  s^2 \La 
   \label{dkas}
\eqy
where $\bar{\ka}= c_2 \ka/c_1$.  
Using \eqref{dotx} we can replace  \eqref{dkas}   by 
\bal
\frac{d x}{d\bar{\ka}}=  -\de x ({\La }  -1/2)\,.
\label{dkax}
\eal
Next, because $H$ is $\ka$ independent, we can use   \eqref{H}  in  \eqref{dkade} to obtain the quadrature
\bq
\frac{d \de}{d \bar\ka}  =   - \frac{e^{1/2-\be_1}}{2} \, \de^2\,  e^{\de^2 \mathring{H}} 
\quad \Rightarrow\quad 
\int^\de_{\mathring{\de}} \frac{e^{-\de'^2 \mathring{H}}}{\de'^2}\,  d\de'=- \frac{e^{1/2-\be_1}}{2}\,\int_{\mathring{\bar\ka}}^{\bar\ka} \!d\bar{\ka}'\,.
\label{quad}
\eq
  With the substitution $u=\de^2\mathring{H}$, \eqref{quad} becomes
\bq
 \int_{\mathring{H}  \mathring{\de}^2}^{\mathring{H}  \de^2}  \frac{e^{-u'}}{u'^{3/2}} \, du'
=-\frac{c_2}{c_1}\,   \frac{e^{1/2-\be_1}}{\sqrt{\mathring{H}}}\, ( \ka -\mathring{\ka})\,.
\label{deka}
\eq
The lefthand side of \eqref{deka} can be written in terms of the   incomplete gamma or error function  
defined by  \eqref{gamaerf}. Then, inverting \eqref{deka} for $\de(\ka)$ and inserting into 
\eqref{dkax} gives the following separable equation:
\bq
\frac{d x}{d\ka} = -\frac{c_2}{c_1}\, x\big(\La (x)-1/2\big) \de(\ka)\,,
\eq
the solution of which yields $x(\ka)$, whence we obtain $s=x\de$ implying  $s(\ka)$.  Finally, upon inserting $s(\ka)$ into \eqref{dotka}, we can obtain $\ka(t)$, and all quantities  are known as functions of time. 




\begin{thebibliography}{10}

\bibitem{MK19}
H.~K. Moffatt and Y.~Kimura.
\newblock Towards a finite-time singularity of the {N}avier-{S}tokes equations
  {P}art 1. {D}erivation and analysis of dynamical system.
\newblock {\em J. Fluid. Mech.}, 861:930--962, 2019.

\bibitem{MK19b}
H.~K. Moffatt and Y.~Kimura.
\newblock Towards a finite-time singularity of the {N}avier-{S}tokes equations
  {P}art 2. {V}ortex reconnection and singularity evasion.
\newblock {\em J. Fluid. Mech.}, 870:R1, 2019.

\bibitem{pjm98}
P.~J. Morrison.
\newblock {H}amiltonian description of the ideal fluid.
\newblock {\em Rev.\ Mod.\ Phys.}, 70:467--521, 1998.

\bibitem{doering}
C.~R. Doering.
\newblock The 3{D} {N}avier-{S}tokes problem.
\newblock {\em Annu. Rev. Fluid Mech.}, 41:109--128, 2009.

\bibitem{B_etal94}
J.~T. Beale, T.~Kato, and A.~Majda.
\newblock Remarks on the breakdown of smooth solutions for the 3{D} {E}uler
  equations.
\newblock {\em Commun. Math. Phys.}, 94:61--66., 1984.

\bibitem{Kt94}
S.~Kida and M.~Takaoka.
\newblock Vortex reconnection.
\newblock {\em Annu. Rev. Fluid Mech.}, 26:169--189, 1994.

\bibitem{K05}
R.~M. Kerr.
\newblock Vortex collapse and turbulence.
\newblock {\em Fluid Dyn. Res.}, 36:249--260, 1994.

\bibitem{HD11}
F.~Hussain and K.~Duraisamy.
\newblock Mechanics of viscous vortex reconnection.
\newblock {\em Phys. Fluids}, 23:021701, 2011.

\bibitem{B_etal08}
G.~P. Bewley, K.~P M.~S.~Paoletti, Sreenivasan, and D.~P. Lathrop.
\newblock Characterization of reconnecting vortices in superfluid helium.
\newblock {\em Proc. Nat. Acad. Sci.}, 105:13707--13710, 2008.

\bibitem{Z_etal12}
S.~Zuccher, M.~Caliari, A.~W. Baggaley, and C.~F. Barenghi.
\newblock Quantum vortex reconnections.
\newblock {\em Phys. Fluids}, 24, 2012.

\bibitem{VPK17}
A.~Villois, D.~Proment, and G.~Krstulovic.
\newblock Universal and non-universal aspects of vortex reconnections in
  superfluids.
\newblock {\em Phys. Rev. F}, 2:125108, 2017.

\bibitem{SP85}
E.~D. Siggia and A.~Pumir.
\newblock Incipient singularities in the {N}aiver-{S}tokes equations.
\newblock {\em Phys. Rev. Lett}, 554:1749--1752, 1985.

\bibitem{PGGB}
R.~B. Pelz, Y.~Gulak Y, J.~M. Greene, and O.~N. Boratav.
\newblock On the finite-time singularity problem in hydrodynamics.
\newblock In A.~Gyr, W.~Kinzelbach, and A.~Tsinober, editors, {\em Fundamental
  Problematic Issues in Turbulence}, Trends in Mathematics, pages 33--40,
  Basel, 1999. Birkhaeuser.

\bibitem{BMH16}
M.~P. Brenner, S.~Hormoz, and A.~Pumir.
\newblock Potential singularity mechanism for the {E}uler equations.
\newblock {\em Phys. Rev. F}, 1:084503, 2016.

\bibitem{KM18}
Y.~Kimura and H.~K. Moffatt.
\newblock A tent model of vortex reconnection under {B}iot-{S}avart evolution.
\newblock {\em J. Fluid Mech.}, 834:R1, 2018.

\bibitem{K18}
R.~M. Kerr.
\newblock Enstrophy and circulation scaling for {N}avier-{S}tokes reconnection.
\newblock {\em J. Fluid Mech.}, 839:R2, 2018.

\bibitem{pjmG80}
P.~J. Morrison and J.~M. Greene.
\newblock Noncanonical {H}amiltonian density formulation of hydrodynamics and
  ideal magnetohydrodynamics.
\newblock {\em Phys. Rev. Lett.}, 45:790--793, 1980.

 \bibitem{s&m}
E.~C.~G.~Sudarshan and N.~Makunda.
\newblock {\em Classical Dynamics:  a modern perspective}.
\newblock John Wiley \& Sons, New York, 1974.

\bibitem{pjmAP20}
P.~J.~Morrison, T.~Andreussi, and F.~Pegoraro.
\newblock 
{L}agrangian and {D}irac constraints for the ideal incompressible fluid and magnetohydrodynamics
\newblock
{\em J. Plasma Phys.}, 86:835860301, 2020.

\bibitem{weinstein}
A.~Weinstein.
\newblock The local structure of {P}oisson manifolds.
\newblock {\em J. Diff. Geom.}, 18:523--557, 1983.
\newblock Erratum { 22}:255, 1985.

\bibitem{pjmY20}
Z.~Yoshida and P.~J. Morrison.
\newblock Deformation of {L}ie-{P}oisson algebras and chirality.
\newblock {\em J. Math. Phys.}, 61:082901, 2020.

\bibitem{a&s}
Milton {Abramowitz} and Irene~A. {Stegun}.
\newblock {\em Handbook of Mathematical Functions with Formulas, Graphs, and
  Mathematical Tables}.
\newblock Dover, New York City, ninth dover printing, tenth gpo printing
  edition, 1964.

\bibitem{pjmMF97}
S.~P. Meacham, P.~J. Morrison, and G.~R. Flierl.
\newblock {H}amiltonian moment reduction for describing vortices in shear.
\newblock {\em Phys.\ Fluids}, 9:2310--2328, 1997.

\bibitem{pjmYT17}
Z.~Yoshida, T.~Tokieda, and P.~J. Morrison.
\newblock Rattleback: A model of how geometric singularity induces dynamic
  chirality.
\newblock {\em Phys. Lett. A}, 381:00, 2017.

\bibitem{c&l}
E.~A. Coddington and N.~Levinson.
\newblock {\em Theory of Ordinary Differential Equations}.
\newblock McGraw-Hill, New York, 1955.

\bibitem{K&R}
E.~A.~Kuznetsov and V.~P.~Ruban.
\newblock 
{H}amiltonian dynamics of vortex lines in hydrodynamic-type systems
\newblock
{\em JETP Letters}, 67:1076--1081, 1998.

\end{thebibliography}
\end{document}